\documentclass{article}

\usepackage{microtype}
\usepackage{graphicx}
\usepackage{subfigure}
\usepackage{booktabs} % for professional tables

\usepackage{hyperref}

\usepackage[accepted]{icml2024}

\usepackage{amsmath}
\usepackage{amssymb}
\usepackage{mathtools}
\usepackage{amsthm}

\usepackage{multicol,multirow}
\usepackage{array}

\usepackage[capitalize,noabbrev]{cleveref}

\theoremstyle{plain}

\theoremstyle{definition}

\theoremstyle{remark}

\usepackage[textsize=tiny]{todonotes}

\icmltitlerunning{Medical Unlearnable Examples}

\begin{document}

\twocolumn[
\icmltitle{Medical Unlearnable Examples: Securing Medical Data from Unauthorized Training via Sparsity-Aware Local Masking}



\icmlsetsymbol{equal}{*}

\begin{icmlauthorlist}
\icmlauthor{Weixiang Sun}{1}
\icmlauthor{Yixin Liu}{2}
\icmlauthor{Zhiling Yan}{2}
\icmlauthor{Kaidi Xu}{3}
\icmlauthor{Lichao Sun}{2}
\end{icmlauthorlist}

\icmlaffiliation{1}{Northeastern University, China}
\icmlaffiliation{2}{Lehigh University}
\icmlaffiliation{3}{Drexel University}

\icmlcorrespondingauthor{Lichao Sun}{lis221@lehigh.edu}

\icmlkeywords{Machine Learning, ICML}

\vskip 0.3in
]



\printAffiliationsAndNotice{\icmlEqualContribution} % otherwise use the standard text.

\begin{abstract}

The rapid expansion of AI in healthcare has led to a surge in medical data generation and storage, boosting medical AI development. However, fears of unauthorized use, like training commercial AI models, hinder researchers from sharing their valuable datasets. To encourage data sharing, one promising solution is to introduce imperceptible noise into the data. This method aims to safeguard the data against unauthorized training by inducing degradation in the generalization ability of the trained model. However, they are not effective and efficient when applied to medical data, mainly due to the ignorance of the sparse nature of medical images. To address this problem, we propose the Sparsity-Aware Local Masking (SALM) method, a novel approach that selectively perturbs significant pixel regions rather than the entire image as previously. This simple yet effective approach, by focusing on local areas, significantly narrows down the search space for disturbances and fully leverages the characteristics of sparsity. Our extensive experiments across various datasets and model architectures demonstrate that SALM effectively prevents unauthorized training of different models and outperforms previous SoTA data protection methods.

\end{abstract}

\section{Introduction}
\label{sec:intro}

The expansion of artificial intelligence (AI) in healthcare has significantly increased the production and storage of sensitive medical data~\cite{rajkomar2018scalable,rasmy2021med}. This data plays a crucial role in advancing research in related fields~\cite{zhang2022shifting,kelly2019key,rasmy2021med}. The more high-quality, open-source datasets that are available, the more contributions can be made by talented researchers to the development of the field~\cite{johnson2019mimic,irvin2019chexpert,pedrosa2019lndb,bilic2023liver,rasmy2021med}. However, many dataset creators are reluctant to open-source their work due to concerns over unauthorized use, such as training commercial models \cite{edwards2022artist,hill2019photos}.

Conventional image classification datasets are relatively easy to obtain and label by laypersons, as shown by ImageNet's use of online crowdsourcing \cite{deng2009imagenet,zhang2021datasetgan}. But in the medical field, annotating is a much more complex process requiring specialized knowledge, usually undertaken by experts like radiologists~\cite{kermany2018identifying,simpson2019large,kather2019predicting}. In addition, it is also necessary to describe the severity of the disease and its relationship with the surrounding tissues. This means that constructing a medical dataset is a labor-intensive task. Unauthorized use could lead to infringement of the creator's rights, leading to a reluctance to release further data publicly. Moreover, from a patient's perspective, concerns about their information being exploited for commercial purposes might decrease their willingness to authorize their data for research~\cite{koh2011data,trinidad2020public}. Therefore, the construction and release of a medical dataset are not only time-consuming and labor-intensive but also fraught with significant ethical and privacy challenges. The reduction in quality open source datasets, resulting from both of the scenarios mentioned, in turn, slows down the development of medical AI~\cite{alberdi2016early,forghani2019radiomics,gunraj2020covidnet}.

To defend unauthorized use and encourage sharing data, \citet{huang2021unlearnable} proposed the technique of contaminating data with imperceptible noise. Models trained on this noise-contaminated data exhibit poor performance for normal utilization. Specifically, these methods create ``unlearnable'' examples from clean data by adding imperceptible noise, demonstrating significant protective capabilities. The design of this error-minimizing noise is based on the intuitive idea that an example with a higher training loss could contain more information to learn. Consequently, this noise protects the data by minimizing the corresponding loss, effectively reducing the informativeness of the data.

Nevertheless, applying this method directly to medical images may not be optimal, as it overlooks the unique properties inherent in medical images~\cite{liu2023securing}. The most important property that is overlooked is sparsity~\cite{ye2012sparse,chuang2007network,huang2009learning,otazo2015low,davoudi2019deep,fang2013fast}. For instance, in cellular microscopy images, even after cropping, there remains a substantial amount of blank background. Similarly, techniques such as CT or tomographic scanning inherently produce sparse data~\cite{chen2018learn,davoudi2019deep,fang2013fast}. Previous methods often struggled to pinpoint specific feature regions in medical data, inadvertently emphasizing sparse areas when generating noise. This led to a major waste of computational resources and suboptimal noise performance, impacts on protection effectiveness, and the time consumed for protection.

To address these challenges, we introduce a novel Sparsity-Aware Local Masking method, which leverages the inherent nature of medical data. Our approach assesses the contribution of pixels to the task based on their gradient and selects pixels with higher contributions for perturbation. This method not only narrows the perturbation search space but also enables the noise generator to focus more on feature regions, yielding noise with stronger protective effects. Additionally, since the protective performance of the noise is entirely derived from feature regions, the performance can be maintained even when cropping large background areas, a common practice in real-world medical workflows. These advancements could significantly motivate medical institutions and researchers to share their data for research or education purposes. In summary, the primary contributions of our research are as follows:

\begin{enumerate}
    \item We are the first to find that the existing Unlearnable Example overlooks the sparse nature of medical data. Specifically, their performance is not optimal and it is not robust against medical-domain data preprocessing.

    \item To address these issues, we propose Sparsity-Aware Local Masking (SALM), specifically designed for the medical domain by limiting the perturbation to features improved protection effectiveness and robustness.

    \item Experiments on multiple medical datasets across different modalities, scales, and tasks demonstrate that our SALM achieves SoTA performance and is consistently effective in various medical scenarios. 
\end{enumerate}

\section{Related Works}

In this paper, we seek to protect medical data from unauthorized exploitation via a data poisoning approach. Data Poisoning is a technique used to compromise the performance of machine learning models on clean data by deliberately altering training samples. This form of attack has proven effective against both Deep Neural Networks (DNNs) and traditional machine learning methods such as SVM~\cite{biggio2012poisoning}. \citet{munoz2017towards} has highlighted the susceptibility of DNNs to data poisoning, although these attacks typically result in only a modest reduction in DNN performance. However, \citet{yang2017generative} found there is a clear distinction between poisoned samples and normal samples, which comes at the cost of reduced data usability. Backdoor Attacks represent a specialized form of data poisoning. Traditional backdoor attacks~\cite{chen2017targeted,liu2020reflection} involve the introduction of falsely labeled training samples embedded with covert triggers into the dataset. A relatively new approach within this realm is the creation of Unlearnable Examples~\cite{huang2021unlearnable}. These are considered a more subtle form of backdoor attack, free of labels and triggers. Unlearnable Examples show promising results in protecting data from unauthorized exploitation in various domains and applications \cite{liu2023unlearnable,liu2023toward,liu2023graphcloak,sun2022coprotector,zhang2023unlearnable,li2023unganable,he2022indiscriminate,zhao2022clpa,salman2023raising,guo2023domain}, such as natural language processing \cite{ji2022unlearnable}, graph learning \cite{liu2023graphcloak}, and contrastive learning \cite{ren2022transferable}. However, the flexibility of protecting medical data has not yet been fully explored. Recently, \citet{liu2023securing} took a first step to evaluate the performance of conventional image protection methods on medical images. Nevertheless, the methods in \cite{liu2023securing} are straightforward adaptations from the previous approach and are suboptimal due to a lack of consideration of the intrinsic characteristics of medical data. To better address this, in this study, we leverage the inherent features of medical data to design more effective protection.

\section{Methodology}

\begin{figure*}[!t]
    \centering
    \includegraphics[width = \linewidth]{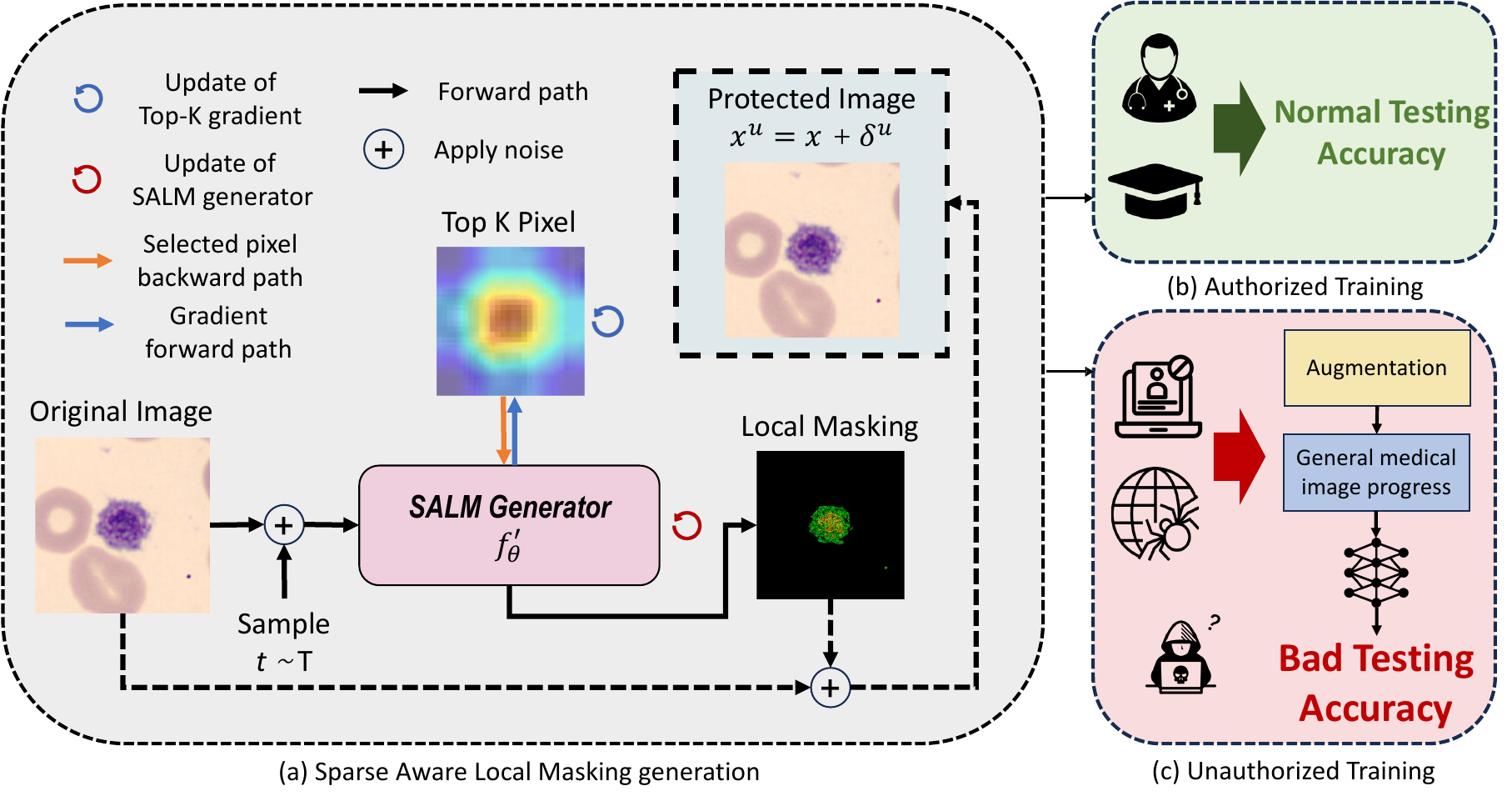}
    \caption{Our SALM method comprises a comprehensive framework that encompasses two primary steps: important pixel acquisition and noise generator training. In the first phase, the model calculates the gradient at each pixel within the image and ranks them, generating a sparse mask through a pre-set $K$ value. In the second phase, the noise generator focuses on perturbing the pixels selected in the previous step and updates its parameters. By implementing this noise, models trained without authorization exhibit poor performance on clean datasets. Conversely, the performance for authorized users remains comparable to that achieved with the original data.}
    \label{fig: framework}
    
\end{figure*}

\subsection{Problem Statement}

\noindent\textbf{Assumptions on Defender’s Capability. } We assume that defenders are capable of making arbitrary modifications to the data they seek to protect, under the premise that these modifications do not impair the visual quality. To better simulate real-world conditions, defenders cannot interfere with the training processes of unauthorized users and do not know the specific models they use. Additionally, once the dataset is publicly released, defenders can no longer modify the data.

\noindent\textbf{Objectives. } This question is posed in the context of utilizing Deep Neural Networks (DNNs) for the classification of medical images. In a classification task comprising $K$ categories, the clean training dataset and test dataset are denoted as $\mathcal{D}_c$ and $\mathcal{D}_t$, respectively.

Suppose the clean training dataset consists of $n$ clean examples, that is, $\mathcal{D}_c=\left\{\left(\boldsymbol{x}_i, y_i\right)\right\}_{i=1}^n$ with $\boldsymbol{x} \in \mathcal{X} \subset \mathbb{R}^d$ are the inputs and $y \in \mathcal{Y}=\{1, \cdots, N\}$ are the labels and $N$ is the total number of classes. We denote its unlearnable version by $\mathcal{D}_u=\left\{\left(\boldsymbol{x}_i^{\prime}, y_i\right)\right\}_{i=1}^n$, where $\boldsymbol{x}^{\prime}=\boldsymbol{x}+\boldsymbol{\delta}$ is the unlearnable version of training example $\boldsymbol{x} \in \mathcal{D}_c$ ,and $\boldsymbol{\delta} \in \Delta \subset \mathbb{R}^d$ is the ``invisible'' noise that makes $\boldsymbol{x}$ unlearnable. The noise $\boldsymbol{\delta}$ is bounded by $\|\delta\|_p \leq \epsilon$ with $\|\cdot\|_p$($L_p$ norm), and $\epsilon$ is set to be small such that it does not affect the normal utility of the example.

In the specific scenario, a DNN model \(f_\theta \) parameterized with $\theta$, is trained on $\mathcal{D}_c$ to learn the mapping from the input domain to the output domain. For simplicity, we will omit the $\theta$ notation in the rest of this paper. To generate an unlearnable dataset, our objective is to induce the model to learn a spurious correlation between noise and labels: \( f_\theta: \Delta \rightarrow \mathcal{Y}, \Delta \neq \mathcal{X} \), when trained on $\mathcal{D}_u$:
\begin{equation}
\label{eq: objective}
    \underset{\theta}{ \min } \mathbb{E}_{\left(\boldsymbol{x}^{\prime}, y\right) \sim \mathcal{D}_u} \mathcal{L}\left(f_\theta\left(\boldsymbol{x}^{\prime}\right), y\right).
\end{equation}

\subsection{Sparsity-Aware Local Masking}

In an ideal scenario, the generation of noise involves a class-matching process to protect each category within \(\mathcal{D}_c\). For simplicity, we define the noise generation process on \(\mathcal{D}_c\) here. Given a clean example \(\boldsymbol{x}\), by generating the imperceptible noise \(\boldsymbol{\Delta}\) for the training input \(\boldsymbol{x}\) by solving the following bi-level optimization problem:

\vspace{-0.7cm}
\begin{equation}
\label{eq: em optim}
    \underset{\theta}{\min } \mathbb{E}_{(\boldsymbol{x}, y) \sim \mathcal{D}_c}\left[\min _{\boldsymbol{\delta}} \mathcal{L}\left(f_\theta^{\prime}(\boldsymbol{x}+\boldsymbol{\delta}), y\right)\right] \ , \text { s.t. }\|\boldsymbol{\delta}\|_p \leq \epsilon,
\end{equation}
where $f^{\prime}$ denotes the source model used for noise generation. Note that this is a min-min bi-level optimization problem: the inner minimization is a constrained optimization problem that finds the $L_p$-norm bounded noise $\delta$ that minimizes the model's classification loss, while the outer minimization problem finds the parameters $\theta$ that also minimize the model's classification loss. However, such an approach does not impose any constraints on the selection of pixels for perturbation. If this method is directly applied to medical data, it overlooks the sparsity inherent in medical datasets.

To bridge the gap in the biomedical domain, motivated by the observation that the important features in the biomedical image are often sparse as we observed before, we propose a sparsity-aware objective that only modifies a portion of important pixels in the image $\boldsymbol{x}$. Formally, we introduce an additional constraint to limit the noise $\boldsymbol{\delta}$ in terms of $\ell_0$ sparsity norm, i.e., $\|\boldsymbol{\delta}\|_0 \leq m$, we have

\vspace{-0.9cm}
\begin{equation}
\label{eq: SALM objective}
    \underset{\theta, \boldsymbol{\delta}:\|\boldsymbol{\delta}\|_p \leq \epsilon \text { and }\|\boldsymbol{\delta}\|_0 \leq m}{ \min } \mathbb{E}_{(\boldsymbol{x}, y) \sim \mathcal{D}_c} \mathcal{L}(f_\theta(\boldsymbol{x}+\boldsymbol{\delta}), y).
\end{equation}

To address the bi-level optimization problem in Eq.~\eqref{eq: em optim}, existing studies have proposed methods such as iterative generator training~\cite{fu2022robust}, target poisoning with a pretrained model~\cite{fowl2021adversarial}. Specifically, in this work, we adopt the iterative generator training framework, with the training termination condition being solely the training steps $M$. When the training step $M$ is sufficient, the noise will be continued optimizing to achieve better performance since there's no accuracy stop condition like~\cite{huang2021unlearnable}. In each step of noise update, we employ the PGD~\cite{madry2017towards} to solve the constrained minimization problem as follows:
% When the training step $M$ is sufficient, due to the absence of an error rate threshold, it is possible to continue optimizing the noise to achieve better performance.
\begin{equation}
    \delta_{t+1}^u=\prod_{\|\delta\|_p \leq \rho_{u}}\left(\delta_{t}^u-\alpha_u \cdot \operatorname{sign}\left(\nabla_{\boldsymbol{x}} \mathcal{L}\left(f_\theta^{\prime}\left(x_{t}+\delta_{t}^u\right), y\right)\right)\right),
\end{equation}

where $t$ is the current step in the training process, $\nabla_{\boldsymbol{x}} \mathcal{L}\left(f_\theta^{\prime}\left(x_{t}^{\prime}\right)\right)$ is the gradient of loss, $\Pi$ is a projection function that clips the noise back to the refined area around the original example $x$ when it goes beyond, $\alpha_u$ is the step size.

Existing methods do not account for the additional \(\ell_0\) constraint as proposed in Eq~\eqref{eq: SALM objective}. In this paper, we leverage the principle that a pixel's contribution to the task is proportional to the magnitude of its gradient. Using the gradient as a basis, we rank pixel importance and subsequently generate a localized mask. This approach aims to achieve both high performance and efficiency. Specifically, the gradient corresponding to \(\boldsymbol{x}\) is defined as \(G_x = \nabla_{\boldsymbol{x}} \mathcal{L}\left(f_\theta^{\prime}\left(\boldsymbol{x}_t^{\prime}\right), y\right)\). The projection \(\mathcal{P}\) is then applied to the obtained gradient:
\begin{equation}
\label{eq: mask}
    \mathcal{P}\left(G_x, k\right)=M \odot G_x, M_{i j}=\left\{\begin{array}{l}
1, g_{i j} \geq g^0(k) \\
0, \text { otherwise }
\end{array}\right.,
\end{equation}

where we seek to modify the top-$k$ percent of the pixels, $g_{i j}$ denotes the value of position $(i, j)$ of the gradient $G_x$, and $g^0(k)$ is the $k$-th percentile value in $G_x$. If $M_{ij}=0$, it means any modification at position $(i, j)$ will result in $\delta_u$ not satisfying the constraint conditions of $\|\boldsymbol{\delta}\|_0 \leq m$.

With projection \(\mathcal{P}\) to refine the local region, the perturbation \(\delta\) is crafted with a PGD~\cite{madry2017towards} process. Given total PGD steps $K_a$, for each iteration $t \in [1,K_a]$ the noise is iteratively updated via:
\begin{equation}
\label{eq: iteration}
    \delta_{t+1}^u=\prod_{\|\delta\|_p \leq \rho_{u}}\left(\delta_{t}^u-\alpha_u \cdot \operatorname{sign}\left(\mathcal{P}\left(g_k,k\right)_{t}, y\right)\right).
\end{equation}

By purposefully reducing the range of perturbation, this approach enables the noise to cover specific feature regions rather than dispersing throughout the entire color space. This ensures that the noise does not expend effort in sparse regions unnecessarily, leading to better performance and efficiency. The overall framework
and procedure are depicted in Figure~\ref{fig: framework} and Appendix~\ref{apx: alg}.

\begin{figure*}[!t]
\centering
    \begin{minipage}{0.33\linewidth}
        \includegraphics[width = \linewidth]{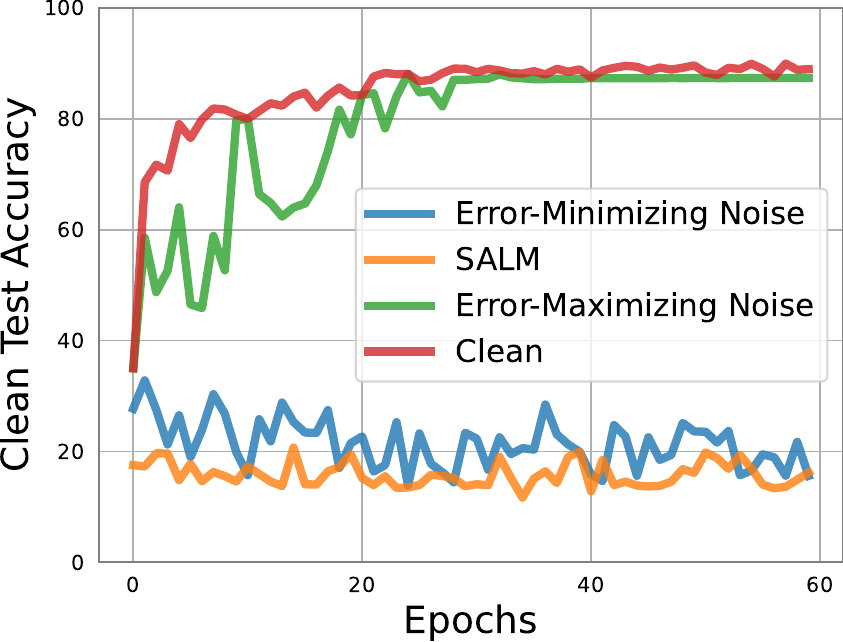}
        \caption{
        The learning curves of ResNet-18 trained on different protected data.
        % Loss Curve. 
        }
        \label{fig: diff method}
    \end{minipage}
    \begin{minipage}{0.3\linewidth}
        \vspace{-0.1cm}
        \includegraphics[width=\linewidth]{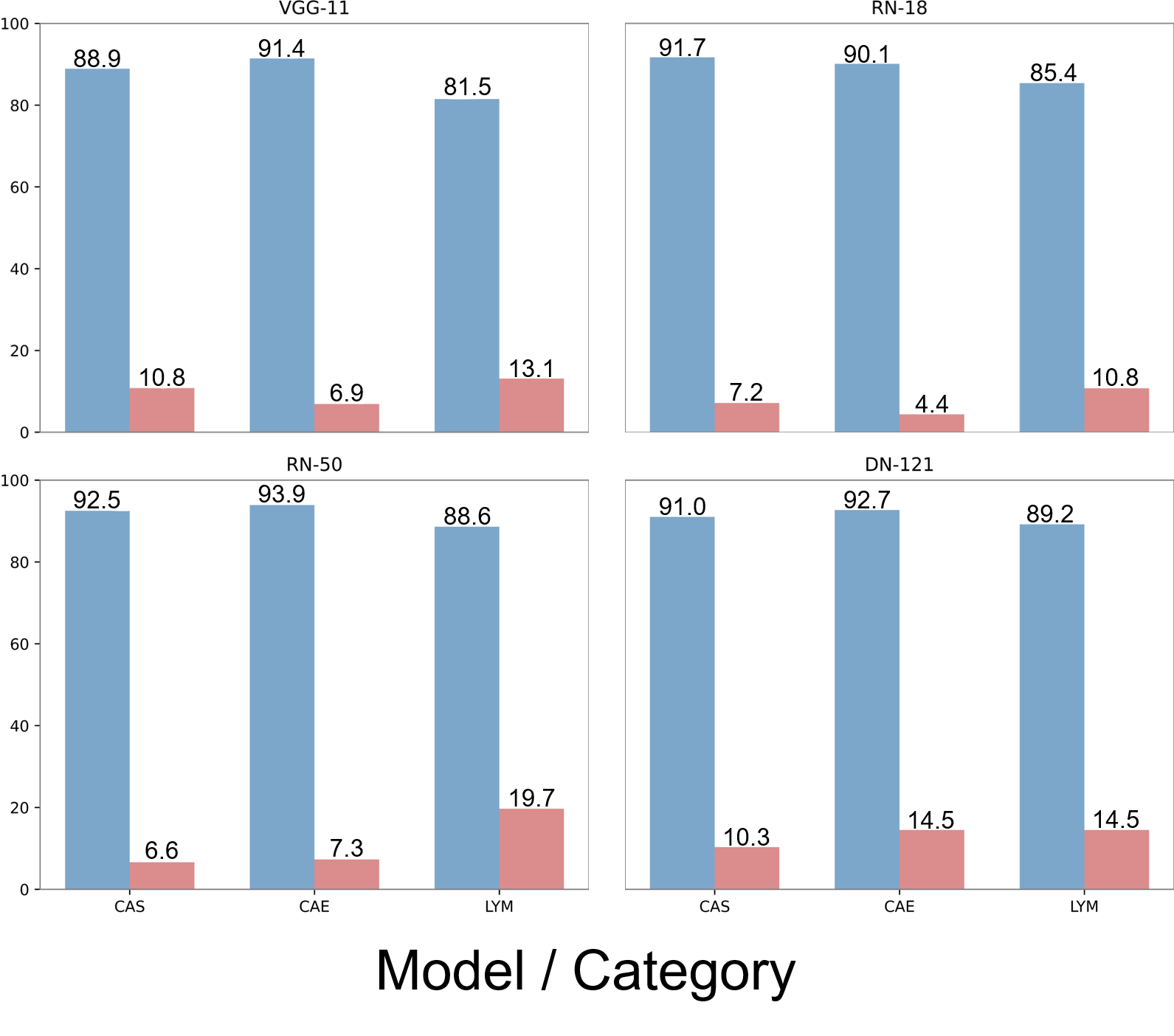}
        \vspace{-0.26cm}
        \caption{
        The selected categories protect effectiveness under different models.
        % Diff. Models. 
        }
        \label{fig: sub}
    \end{minipage}
    \begin{minipage}{0.33\linewidth}
        \includegraphics[width = \linewidth]{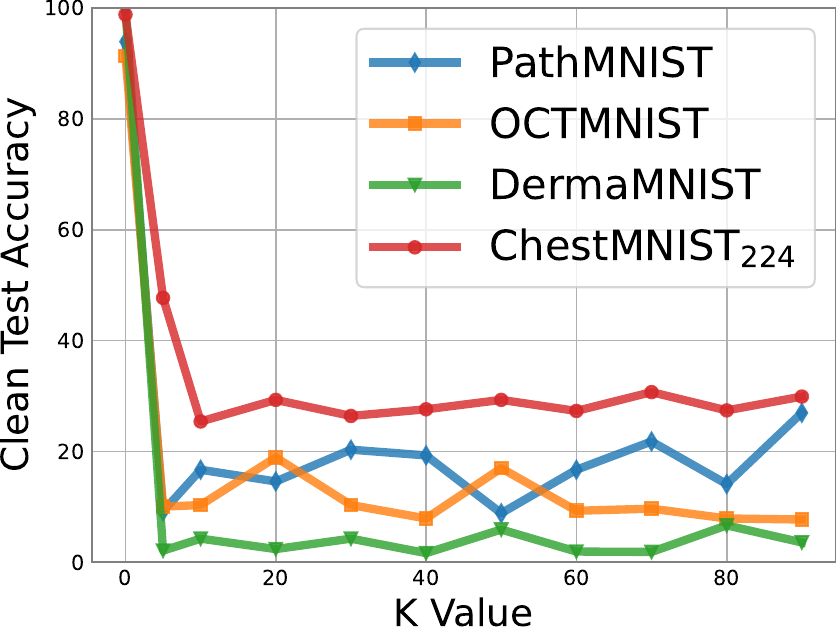}
        \caption{The effect of $K$ on clean test accuracy(\%) for the four datasets.}
        \label{fig: diff k}
    \end{minipage}
    
    % \caption{\textbf{Left:} The learning curves of ResNet-18 trained on different protected data. \textbf{Right: }The effect of different $K$ values on clean test accuracy(\%) for the four datasets.}

\end{figure*}

\section{Experiments and Results}

We selected more than 14 datasets from Medmnist~\cite{medmnistv1} and Medmnist-v2~\cite{medmnistv2} and conducted extensive experiments on various models following \cite{huang2021unlearnable}. 
% Additionally, we chose three excellent conventional image preservation algorithms as baselines. 
We compare SALM with three baselines, including \textbf{AdvT} \cite{fowl2021adversarial}, \textbf{EM} \cite{huang2021unlearnable}, \textbf{SP} \cite{yu2022availability}.
For more experimental details, please refer to Appendix~\ref{apx: exp}. More experiments, analysis, and the case study can be found in Appendix~\ref{apx: more} and \ref{apx: case}.

\begin{table}[!t]
    \centering
        \caption{Clean test accuracy (\%) of RN-18 trained on datasets across various modalities protected by different methods. The symbol $\downarrow$ in the following context indicates a decrease in accuracy compared to the clean test accuracy. }
    \resizebox{\linewidth}{!}{
    \begin{tabular}{c|c|c|c|c|c}
    \hline
         \textbf{Dataset} & \textbf{Clean} & \textbf{AdvT} & \textbf{EM} & \textbf{SP} & \textbf{SALM($K=10$)} \\ \hline
         PathMNIST & 90.7 & \textbf{8.4($\downarrow$82.3)} & 13.5($\downarrow$77.2) & 12.2($\downarrow$78.5) & 11.8($\downarrow$78.9) \\
         ChestMNIST & 94.7 & 38.8($\downarrow$55.9) & 78.7($\downarrow$16.0) & 63.5($\downarrow$31.2) & \textbf{27.6($\downarrow$67.1)} \\
         DermaMNIST & 73.5 & 38.7($\downarrow$34.8) & 17.1($\downarrow$56.4) & 11.5($\downarrow$62.0) & \textbf{8.7($\downarrow$64.8)}\\
         OCTMNIST & 74.3 & 21.4($\downarrow$52.9) & 25.0($\downarrow$49.3) & 22.7($\downarrow$51.5) & \textbf{21.2($\downarrow$53.1)} \\
         PneumoniaMNIST & 85.4 & 66.9($\downarrow$18.5) & 65.8($\downarrow$19.6) & 70.1($\downarrow$15.3) & \textbf{63.9($\downarrow$21.5)} \\
         RetinaMNIST & 52.4 & 39.5($\downarrow$12.9) & 13.0($\downarrow$39.4) & \textbf{8.3($\downarrow$44.1)} & 12.2($\downarrow$40.2) \\
         BreastMNIST & 86.3 & \textbf{44.9($\downarrow$41.4)} & 46.2($\downarrow$40.1) & 50.0($\downarrow$36.3) & 53.2($\downarrow$33.1) \\
         BloodMNIST & 95.8 & 19.6($\downarrow$76.2) & 30.5($\downarrow$65.3) & 30.5($\downarrow$65.3) & \textbf{7.1($\downarrow$88.7)} \\
         TissueMNIST & 67.6 & 19.7($\downarrow$47.9) & 17.4($\downarrow$50.2) & \textbf{7.3($\downarrow$60.3)} & 21.4($\downarrow$46.2) \\
         OrganAMNIST & 93.5 & 81.0($\downarrow$12.5) & 78.1($\downarrow$15.4) & 86.1($\downarrow$7.4) & \textbf{69.0($\downarrow$24.5)} \\
         OrganCMNIST & 90.0 & 70.3($\downarrow$19.7) & 72.2($\downarrow$17.8) & 74.8($\downarrow$15.2) & \textbf{64.6($\downarrow$25.4)} \\
         OrganSMNIST & 78.2 & 51.7($\downarrow$26.5) & 50.1($\downarrow$28.1) & 53.0($\downarrow$25.2) & \textbf{46.8($\downarrow$31.4)} \\
         PathMNIST$_{224}$  & 90.9 & 70.8($\downarrow$21.9) & 32.5($\downarrow$60.2) & 29.1($\downarrow$63.6) & \textbf{27.5($\downarrow$67.3)} \\ 
         ChestMNIST$_{224}$  & 92.7 & 31.6($\downarrow$61.1) & 64.4($\downarrow$28.3) & 50.1($\downarrow$42.6) & \textbf{25.4($\downarrow$67.3)} \\ 
         
         \hline
    \end{tabular}}

    \label{tab: compare}
\end{table}

\textbf{Effectiveness Analysis. }We initially selected PathMNIST for comparison between our method and the conceptually similar error-minimizing and error-maximizing noise. The results in Figure~\ref{fig: diff method} show that from the onset of training, both our method and the error-minimizing noise provide effective protection, with our method demonstrating superior performance. Conversely, the efficacy of error-maximizing noise gradually diminished throughout the training. To further evaluate the performance of our method, we directly transferred several conventional image protection methods to medical and compared them with our approach. The results in Table~\ref{tab: compare} confirm that our method is not only broadly applicable to different modalities of medical datasets but also outperforms the previous SoTA methods.

% \subsection{Stability under Different Arch}
\begin{table}[t]
\centering
\caption{Clean test accuracy (\%) of DNNs trained on the clean training sets ($\mathcal{D}_{c}$) or their unlearnable ones ($\mathcal{D}_{u}$) made by different $K$ value.}
\resizebox{\linewidth}{!}{
\begin{tabular}{c|c|p{0.94cm}<{\centering}c|p{1.1cm}<{\centering}c|p{0.83cm}<{\centering}c|p{1.22cm}<{\centering}c}
\hline
\multirow{2}{*}{\textbf{$K$}} & \multirow{2}{*}{\textbf{Model}} & \multicolumn{2}{c|}{\textbf{PathMNIST}} & \multicolumn{2}{c|}{\textbf{DermaMNIST}} & \multicolumn{2}{c|}{\textbf{OctMNIST}} & \multicolumn{2}{c}{\textbf{ChestMNIST$_{224}$}} \\
          &  & \textbf{$\mathcal{D}_{c}$} & \textbf{$\mathcal{D}_{u}$} & \textbf{$\mathcal{D}_{c}$} & \textbf{$\mathcal{D}_{u}$} & \textbf{$\mathcal{D}_{c}$} & \textbf{$\mathcal{D}_{u}$} & \textbf{$\mathcal{D}_{c}$} & \textbf{$\mathcal{D}_{u}$} \\
          \hline
 
\multirow{4}{*}{$K=10$} & VGG-11 & 91.7 & \textbf{19.2}  & 73.9 & \textbf{10.9} & 78.2 & \textbf{21.4} & 91.7 & \textbf{27.3} \\
 & RN-18 & 90.7 & \textbf{11.8}  & 73.5 & \textbf{8.7} & 74.3 & \textbf{21.2} & 92.7 & \textbf{25.4}\\
 & RN-50 & 94.6 & \textbf{21.9}  & 74.6 & \textbf{9.8} & 74.3 & \textbf{19.0} & 94.7 & \textbf{25.4}\\
 & DN-121 & 96.4 & \textbf{14.8} & 75.3 & \textbf{14.5} & 75.6 & \textbf{13.3} & 95.1 & \textbf{21.2}\\ \hline

 \multirow{4}{*}{$K=30$} & VGG-11 & 90.8 & \textbf{18.3} & 74.4 & \textbf{11.8} & 75.5 & \textbf{15.1} & 92.1 & \textbf{31.4}\\
 & RN-18 & 92.9 & \textbf{14.1} & 73.5 & \textbf{4.0} & 74.4 & \textbf{12.2} & 94.6 & \textbf{24.0}\\
 & RN-50 & 94.2 & \textbf{17.4} & 74.4 & \textbf{9.3} & 77.0 & \textbf{14.3} & 94.8 & \textbf{30.7}\\
 & DN-121 & 94.2 & \textbf{19.7} & 77.2 & \textbf{11.8} & 74.9 & \textbf{30.0} & 94.5 & \textbf{29.9}\\ \hline
\end{tabular}
}

\label{tab: diff arch}

\end{table}

\noindent\textbf{Different Architectures Selection. }Before releasing datasets, we cannot foresee the specific training model unauthorized users might adopt. Hence, ensuring that the protected data remains effective across various models is crucial. While protectors can only select a single source model for noise generation, unauthorized users are free to choose any model, seemingly placing the protectors at a disadvantage place. However, in reality, the choice of model by unauthorized users does not impact the efficacy of our method. The results in Table~\ref{tab: diff arch} demonstrate that the SALM with the source model ResNet-18~\cite{he2016deep} is effective across various models. The results in Figure~\ref{fig: sub} show the randomly extracted subset still has excellent performance under different models. Furthermore, the protection performance is not necessarily best when unauthorized trained on ResNet-18. This indicates that our method is not limited by the source model and the target model, enabling more effective application in real-world scenarios and facilitating the open-sourcing of high-quality datasets. What's more, the protective effect when $K$ is 30 is not significantly better than when it is 10. This prompts us to delve deeper into the impact of the choice of $K$-value.

\noindent\textbf{Different $K$ Selection. }In this section, we delve deeper into our approach to unveil the nuances of our method's dependency on the choice of $K$ value. $K$ values were selected within a range from 0 to 90 in increments of ten. We also tried $K=5$ to test the limitation of SALM. When $K$ is set to 0, it implies that the model is trained on clean data. The results in Figure~\ref{fig: diff k} show when $K$ is set at 5 or higher, the noise proves to be generally effective. Consequently, in subsequent sections, unless specifically emphasized, we opt for $K=10$ for our experiments. This underscores the effectiveness of limiting the perturbation search space and also demonstrates that the existing methods are not optimal in the medical domain. What's more, it highlights the high efficiency of our approach to medical data, which can enable the accelerated release of datasets.

% \subsection{Resistance to Low-pass Filters}
\begin{table}[!t]
    \caption{Clean test accuracy (\%) of RN-18 trained on datasets protected by different methods after three common low-pass filters.}
    \label{tab: fitr}
    \centering
    \resizebox{\linewidth}{!}{
    \begin{tabular}{c|c|c|c|c|c|c}
    \hline
    \textbf{Dataset} & \textbf{Filter} & \textbf{Clean} & \textbf{AdvT} & \textbf{EM} & \textbf{SP} & \textbf{SALM} \\
    \hline
    \multirow{3}{*}{PathMNIST} & Mean & 91.3 & 17.9($\downarrow$73.4) & 16.4($\downarrow$74.9) & 17.1($\downarrow$74.2) & \textbf{15.6($\downarrow$75.7)} \\
         & Median & 90.8 & 22.4($\downarrow$68.4) & \textbf{7.2($\downarrow$83.6)} & 12.7($\downarrow$78.1) & 19.3($\downarrow$71.5) \\
         & Gaussian & 92.1 & 35.5($\downarrow$56.6) & \textbf{11.0($\downarrow$81.1)} & 18.0($\downarrow$74.1) & 17.7($\downarrow$74.4) \\
    \hline
    \multirow{3}{*}{DermaMNIST} & Mean & 73.8 & 45.8($\downarrow$28.0) & 31.9($\downarrow$41.9) & 23.9($\downarrow$49.9) & \textbf{21.5($\downarrow$52.3)} \\
         & Median & 75.1 & 62.6($\downarrow$12.5) & 34.6($\downarrow$40.5) & 19.9($\downarrow$55.2) & \textbf{17.8($\downarrow$57.3)} \\
         & Gaussian & 74.7 & 40.1($\downarrow$34.6) & 33.1($\downarrow$41.6) & \textbf{25.8($\downarrow$48.9)} & 40.6($\downarrow$34.1) \\
    \hline
    \multirow{3}{*}{OctMNIST} & Mean & 77.7 & 25.7($\downarrow$52.0) & 28.3($\downarrow$49.4) & \textbf{23.8($\downarrow$53.9)} & 29.6($\downarrow$48.1) \\
         & Median & 79.6 & 23.5($\downarrow$56.1) & 25.7($\downarrow$53.9) & 23.6($\downarrow$56.0) & \textbf{22.6($\downarrow$57.0)} \\
         & Gaussian & 77.7 & \textbf{22.9($\downarrow$54.8)} & 26.1($\downarrow$51.6) & 23.1($\downarrow$54.6) & 29.2($\downarrow$48.5) \\
    \hline
    \multirow{3}{*}{ChestMNIST$_{224}$} & Mean & 91.4 & 37.7($\downarrow$53.7) & 72.5($\downarrow$18.9) & 58.2($\downarrow$33.2) & \textbf{27.0($\downarrow$64.4)} \\
         & Median & 90.9 & 38.2($\downarrow$52.7) & 62.7($\downarrow$28.2) & 50.9($\downarrow$40.0) & \textbf{28.4($\downarrow$62.5)} \\
         & Gaussian & 91.1 & 45.1($\downarrow$46.0) & 70.8($\downarrow$20.3) & 51.4($\downarrow$39.7) & \textbf{29.8($\downarrow$61.3)} \\
    \hline

    \end{tabular}}
    
\end{table}
\noindent\textbf{Resistance to Low-pass Filters. }
Constrained by limitations in the quality of medical imaging, the acquired images often exhibit noise. Consequently, researchers commonly employ low-pass filters for pre-processing due to their simplicity and efficiency. Therefore, verifying the resilience of our method to such filtering is of paramount importance. We chose three low-pass filters: Mean, Median, and Gaussian, each with a 3$\times$3 window size. The results in Table~\ref{tab: fitr} show that our method retains its effectiveness post-filtering, and the sensibility is lower than other methods, affirming its applicability within actual medical workflows.

\section{Conclusion}
This work introduces the SALM method, a novel approach dedicated to generating Unlearnable Examples specifically designed for medical datasets. Extensive observation has revealed that medical datasets are inherently sparse, a characteristic not effectively utilized by existing methods for generating Unlearnable Examples. Consequently, the SALM method is designed to selectively perturb only a specific subset of critical pixels. Extensive experimental results demonstrate that the SALM method effectively protects medical images from unauthorized training. Simultaneously, it ensures stability and effectiveness throughout common medical image processing workflows (e.g., filtering and cropping). Additionally, the processed images retain their visual usability, not impeding clinical diagnosis by physicians. Furthermore, we demonstrate that SALM has strong, flexible applicability in practical application scenarios.

\nocite{langley00}

\bibliography{main}
\bibliographystyle{icml2024}

%%%%%%%%%%%%%%%%%%%%%%%%%%%%%%%%%%%%%%%%%%%%%%%%%%%%%%%%%%%%%%%%%%%%%%%%%%%%%%%
%%%%%%%%%%%%%%%%%%%%%%%%%%%%%%%%%%%%%%%%%%%%%%%%%%%%%%%%%%%%%%%%%%%%%%%%%%%%%%%
% APPENDIX
%%%%%%%%%%%%%%%%%%%%%%%%%%%%%%%%%%%%%%%%%%%%%%%%%%%%%%%%%%%%%%%%%%%%%%%%%%%%%%%
%%%%%%%%%%%%%%%%%%%%%%%%%%%%%%%%%%%%%%%%%%%%%%%%%%%%%%%%%%%%%%%%%%%%%%%%%%%%%%%
\newpage
\appendix
\onecolumn
\section{More Related Work}
\textbf{Medical Data Protection. }As information technology advances, digital technologies are increasingly integrated into medicine, affecting both clinical treatment and scientific research~\cite{senbekov2020recent,rajpurkar2017mura,johnson2019mimic,irvin2019chexpert,pedrosa2019lndb,bilic2023liver,rasmy2021med,zhang2022shifting,kelly2019key}. However, these developments also present significant risks and challenges regarding patient privacy, medical information breaches occur frequently around the world~\cite{alpert2003protecting,kayaalp2018patient,price2019privacy,murdoch2021privacy}. Inadequate data protection can lead to substantial harm through the leakage of personal information, such as the disclosure of sensitive health conditions like HIV, potentially leading to social isolation and psychological disorders~\cite{gostin2000public}. Furthermore, incidents of data breaches may also reduce patients' trust in medical research institutions, making them reluctant to share their data~\cite{koh2011data,trinidad2020public}. Earlier studies focused on robust physical protection measures, like using encrypted storage devices~\cite{heurix2011privacy,akinyele2011securing,sun2020blockchain}, establishing firewalls~\cite{barrows1996privacy,liu2012enhancement}, and secure communication transmission modes~\cite{gong2015medical,li2016secure}, to safeguard data. \citet{liu2018clustering} introduced a clustering method based on $K$-anonymity algorithm as the building block of privacy-preserving for medical devices' data. However, as collaboration among various research institutions intensifies and open-source data sharing on the internet becomes more crucial, these methods become less applicable. \citet{baowaly2019synthesizing,yoon2020anonymization} attempted to use generated data to reduce granularity and thus protect privacy, but this method involves a significant trade-off between information loss and protection efficacy. However, our method remains robust under various conditions during the construction process of real-world datasets and effectively balances high data usability with consistent protection.

\section{SALM Algorithm}
\label{apx: alg}
\begin{algorithm}[h]
   \caption{Training SALM generator and generating noise.}
   \label{alg: salm}
\begin{algorithmic}
   \STATE {\bfseries Input:} Training data set $\mathcal{T}$, Training steps $M$, PGD parameters $\alpha_u$ and $\rho_u$, transformation distribution $T$, the sampling number $J$ for gradient approximation.
   \STATE {\bfseries Initialization:} source model parameter $\theta$, $\delta^u$.
   \STATE // Following Eq.~\eqref{eq: SALM objective}
   \FOR{$i$ \textbf{in} $1, \cdots, M$}
            \STATE Sample minibatch $(x, y) \sim \mathcal{T}$, sample transformation $t_j \sim T$
            \STATE Calculate $g_k \leftarrow \frac{1}{J} \sum_{j=1}^J \frac{\partial}{\partial \delta^u} \ell(f'_\theta(t_j(x+\delta^u) ), y)$
            \STATE Determine $k$-th percentile value $g^{0}(k)$ in $g_k$
            \FOR{each element $g_{ij}$ in $g_k$}
                \IF{{$g_{ij} \ge g^{0}(k)$}}
                    \STATE Set $M_{ij} = 1$
                \ELSE
                    \STATE Set $M_{ij} = 0$
                \ENDIF
            \ENDFOR
            \STATE Apply local mask: $\mathcal{P}(g_k, k) = M \odot g_k$ \quad // Following Eq.~\eqref{eq: mask}
            \STATE Update $\delta^u \leftarrow \prod_{\|\delta\|\leq\rho_u} \left( \delta^u - \alpha_u \cdot \mathrm{sign}(\mathcal{P}(g_k, k)) \right)$
            \STATE Update source model parameter $\theta$ based on minibatch $(t(x+\delta^u),y)$
        \ENDFOR
        
   \STATE {\bfseries Output: SALM noise generator $f'_\theta$, SALM noise $\delta^u$}
\end{algorithmic}
\end{algorithm}

\section{Experiments Setup}
\label{apx: exp}
\textbf{Datasets. }Our research involved extensive experiments on MedMNIST~\cite{medmnistv1}, which is a comprehensive collection of standardized biomedical images spanning 12 distinct datasets. MedMNIST includes the major modalities in medical imaging.
%, including OCT (Optical coherence tomography), X-ray, and Microscope, among others. 
This dataset showcases diverse data scales and a range of tasks.
%, including binary/multi-class classification, ordinal regression, and multi-label tasks. 
Furthermore, to more effectively assess our method's performance on large, real-world datasets, we also do experiments on the 224x224 datasets from MedMNIST-v2~\cite{medmnistv2}.

\textbf{Model and Implementation Details. }We selected the following well-known models to evaluate our method against benchmark approaches: VGG-11~\cite{simonyan2014very}, ResNet-18/50~\cite{he2016deep} and DenseNet-121~\cite{huang2017densely}. For the SALM generator, we select RN-18 as the source model to generate noise. We set the perturbation noise $\rho_u=8/255$ as default. We choose an SGD optimizer for both noise generation and training. Both of their weight decay are set to 5e-4, and momentum is set to 0.9. The initial learning rate is set to 0.1 with a decay rate of 0.1.

\textbf{Baselines. }Since there's no previous method specifically designed for medical data, we compare our Sparsity-Aware Local Masking \textbf{(SALM)} with the existing SoTA methods in the general image domain. The baseline methods are as following: \textbf{T}arget \textbf{A}dversarial \textbf{P}oisoning \textbf{(TAP)}~\cite{fowl2021adversarial}, \textbf{E}rror-\textbf{M}inimizing noise \textbf{(EM)}~\cite{huang2021unlearnable}, \textbf{S}ynthesized \textbf{P}erturbation \textbf{(SP)}~\cite{yu2022availability}. None of these three methods impose restrictions on the range of pixels to be perturbed. TAP and EM both target the training loss and craft noise using the gradient information to trigger maximum error or trick models to overfit. SC, on the other hand, imposes hand-crafted linear-separable noise into the data, thereby leading the model to learn only simple noise-label correlation.

\section{More Experiments}

\begin{table}[!t]
    \centering
        \caption{More results of RN-18 trained on 224 version datasets protected by different methods. The symbol $\downarrow$ in the following context indicates a decrease in accuracy compared to the clean test accuracy. }
    \resizebox{\linewidth}{!}{
    \begin{tabular}{c|c|c|c|c|c}
    \hline
         \textbf{Dataset} & \textbf{Clean} & \textbf{AdvT} & \textbf{EM} & \textbf{SP} & \textbf{SALM($K=10$)} \\ \hline

         DermaMNIST$_{224}$ & 74.8 & 19.2($\downarrow$55.6) & 11.4($\downarrow$63.4) & 10.0($\downarrow$64.8) & \textbf{9.3($\downarrow$65.5)}\\
         OCTMNIST$_{224}$ & 76.2 & 32.3($\downarrow$43.9) & \textbf{14.6($\downarrow$61.6)} & 20.6($\downarrow$55.6) & 25.8($\downarrow$50.4) \\
         PneumoniaMNIST$_{224}$ & 84.1 & 66.7($\downarrow$17.4) & 67.2($\downarrow$16.9) & 61.6($\downarrow$22.5) & \textbf{57.4($\downarrow$26.7)} \\
         RetinaMNIST$_{224}$ & 51.0 & 27.6($\downarrow$23.4) & 21.1($\downarrow$29.9) & 10.5($\downarrow$40.5) & \textbf{9.1($\downarrow$41.9)} \\
         BreastMNIST$_{224}$ & 80.1 & \textbf{35.1($\downarrow$45.0)} & 40.0($\downarrow$40.1) & 40.2($\downarrow$39.9) & 37.8($\downarrow$42.3) \\
         BloodMNIST$_{224}$ & 96.0 & 16.3($\downarrow$80.3) & 16.6($\downarrow$79.4) & 7.0($\downarrow$89.0) & \textbf{6.3($\downarrow$89.7)} \\
         TissueMNIST$_{224}$ & 68.4 & 18.5($\downarrow$49.9) & 23.6($\downarrow$44.8) & \textbf{16.7($\downarrow$51.7)} & 19.1($\downarrow$49.3) \\
         OrganAMNIST$_{224}$ & 95.2 & 85.8($\downarrow$9.4) & 83.5($\downarrow$11.7) & 85.3($\downarrow$9.9) & \textbf{82.1($\downarrow$13.1)} \\
         OrganCMNIST$_{224}$ & 91.6 & 82.3($\downarrow$9.3) & 69.7($\downarrow$21.9) & 72.2($\downarrow$19.4) & \textbf{69.3($\downarrow$22.3)} \\
         OrganSMNIST$_{224}$ & 77.8 & 47.8($\downarrow$30.0) & 44.6($\downarrow$33.2) & 65.2($\downarrow$12.6) & \textbf{42.1($\downarrow$35.7)} \\
         
         \hline
    \end{tabular}}

    \label{tab: sup compare}
\end{table}

\label{apx: more}

\begin{table}[t]
    \centering
    \begin{minipage}[b]{0.53\textwidth}
        \centering
        \caption{Test accuracy (\%) of RN-18 trained on different kinds of data of PathMNIST.}
        \label{tab: crop}
        \resizebox{\linewidth}{!}{
        \begin{tabular}{cccccc}
         \toprule
         \textbf{Method} & $\mathcal{D}_c$ & $\text{Cropped}( \mathcal{D}_{c})$ & $\mathcal{D}_u$ & $\text{Cropped}( \mathcal{D}_{u})$ & Gap \\
         \midrule
         TAP & 91.2 & 81.9 & 26.5  & 12.7 & 4.5 $\downarrow$\\
         EM & 90.6 & 82.2 & 17.2 & 4.7 & 4.1 $\downarrow$\\
         SP & 90.9 & 81.6 & 21.3 & 12.3 & 0.2 $\uparrow$\\
         SALM & 90.9 & 82.5 & 14.9 & 5.3 & \textbf{1.2}\textbf{$\uparrow$} \\
         
        \bottomrule 
        \end{tabular}
        }
    \end{minipage}
\hspace{1mm}
    \begin{minipage}[b]{0.43\textwidth}
        \centering
        \caption{Image similarity scores of different methods.}
        \label{tab: eva}
        
        \resizebox{\linewidth}{!}{
        \begin{tabular}{cccccc}
         \toprule
         \textbf{Method} & SSIM & aHash & pHash & dHash & NMI  \\
         \midrule
         TAP & 94.6 & 75.1 & 88.8 & 77.2 & 42.7\\
         EM & 92.7 & 77.7 & 89.1 & 77.4 & 42.9 \\
         SP & 99.9 & 95.4 & 99.9 & 99.3 & 99.9 \\
         SALM & 94.4 & 77.9 & 89.4 & 77.2 & 44.7 \\
         
        \bottomrule 
        \end{tabular}

        }
    \end{minipage}
    
\end{table}
\subsection{Resistance to Cropping. }As we consistently highlighted, medical data exhibits greater sparsity compared to conventional data. For instance, when processing microscopic images, physicians or researchers typically reduce the proportion of the background through cropping and magnifying details to facilitate analysis. In this section, we further validate the effectiveness of our SALM after cropping. We calculate the difference in test accuracy on the protected data before and after cropping. To minimize the impact of cropping itself on performance, we add the difference in test accuracy on the clean data before and after cropping, which we denote as the Gap value. The results in Table~\ref{tab: crop} show after cropping, the performance gap of TAP and EM significantly narrows. This finding confirms that color space disturbances by these methods are random. If perturbations are concentrated in non-feature areas like transparent slides, cropping might result in a loss of protection. Conversely, SALM focuses on pixels that significantly contribute to the model, which ensures that removing the background does not impact the regions with more protective noise and thus preserves the protection.

\subsection{Similarity Compare. }Although previous studies~\cite{liu2023securing,huang2021unlearnable,fu2022robust,fowl2021adversarial,yu2022availability} have demonstrated that perturbation radii as small as 8 or even 16 do not affect human visual perception, considering the sensitivity of medical images, we still assessed the similarity. Specifically, we utilized the Structural Similarity Index Measure (SSIM), which assesses luminance, contrast, and structure, along with various Hash methods focusing on low-frequency information. The results in Table~\ref{tab: eva} reveal that SYN has the highest scores. This is likely due to its structure, consisting of square regions of differing colors with minimal distinctions within these areas. However, compared with other methods, despite being based on feature perturbations, our method does not substantially affect images. This underscores our approach's ability to effectively preserve image utility.

\begin{figure}[!t]
    \centering
    \includegraphics[width = \linewidth]{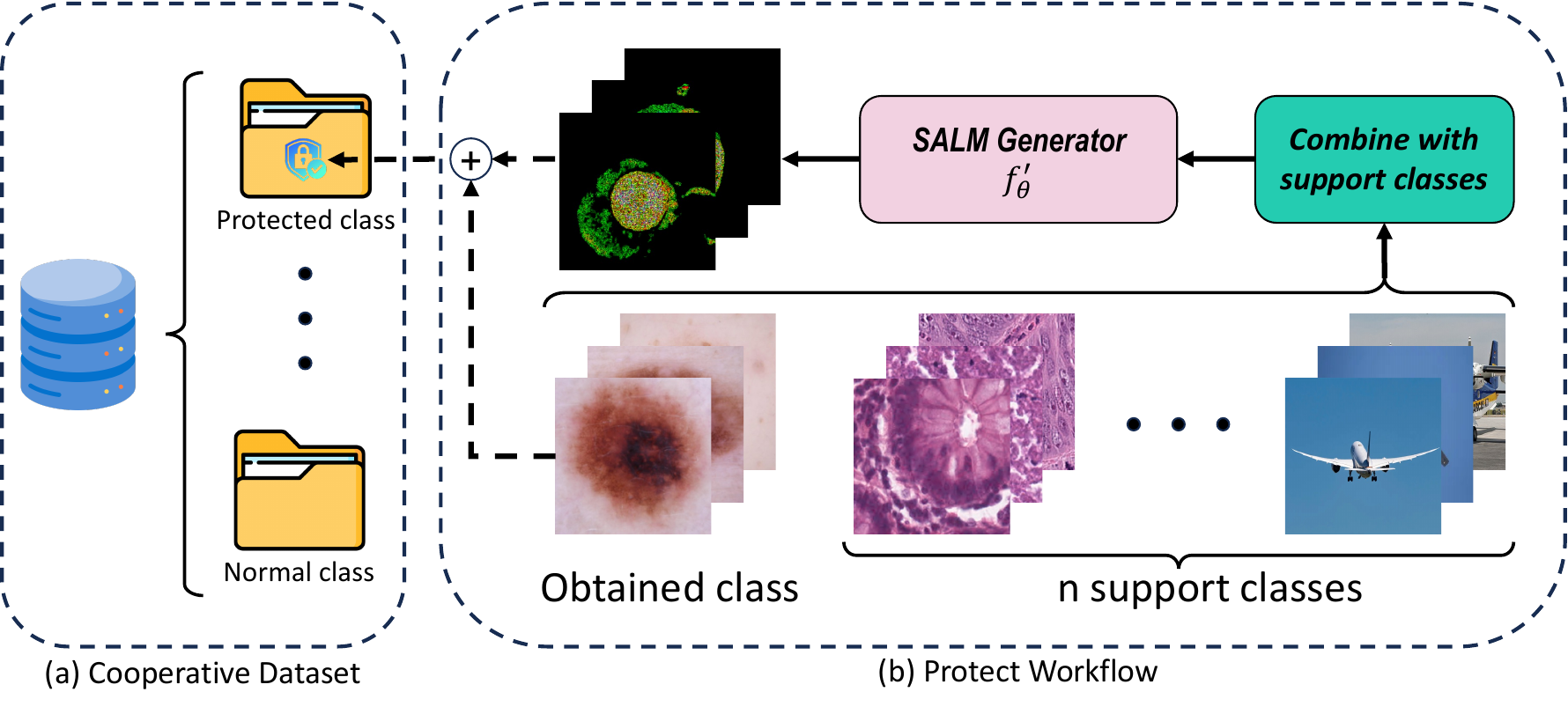}
    \caption{The protected framework in the case of the class-combined dataset. The noise generated for the corresponding obtained class from data combined with other support classes still retains its protective effect.}
    \label{fig: case study}
\end{figure}

\section{Case Study}
\label{apx: case}

\subsection{Case 1: Class-wise Combined Dataset}

We initially conducted a study on the scenario of the class-wise combination. To better explain the experimental setting, we introduce the concept of ``support classes,'' which is utilized to augment the existing single-class data. We selected two kinds of support classes: medical classes and general classes. The framework is shown in Figure~\ref{fig: case study}. First, we chose melanoma from the DermaMNIST as a protected target and selected all classes within the PathMNIST as support classes for our experiments. The melanoma data was combined with PathMNIST to create a ten-class dataset while only preserving the SALM noise generated for the melanoma. Subsequently, to broaden the selection range of support classes, we also selected some general image classes (e.g., airplanes) as support classes.
% for our experiments.

Following the generation of class-specific noise, we utilized the same experimental settings as in previous tests. Using PathMNIST for support classes, the generated noise resulted in a clean test accuracy of \textbf{9.9\%}. Noise generated with general image classes from ImageNet~\cite{deng2009imagenet} as support classes led to an accuracy of \textbf{8.1\%}. The choice of either type of support class does not affect the noise performance. This demonstrates the flexible applicability of our method.

\subsection{Case 2: Sample-wise Combined Dataset}
\begin{table}[!t]
\centering
\caption{Effectiveness under different unlearnable percentages on PathMNIST with RN-18 model: lower clean accuracy indicates better effectiveness. \textbf{$\mathcal{D}_{u}+\mathcal{D}_{c}$}: a mix of unlearnable and clean data; \textbf{$\mathcal{D}_{c}$}: only the clean proportion of data. Percentage of unlearnable examples: $\frac{\mathcal{D}_u}{\mathcal{D}_c+\mathcal{D}_u}$.}
\label{tab: diff per}
\resizebox{\linewidth}{!}{
\begin{tabular}{c|clccccccccc}
\hline
\multirow{3}{*}{\textbf{\begin{tabular}[c]{@{}c@{}}$K$\\ Value\end{tabular}}} & \multicolumn{11}{c}{\textbf{Percentage of unlearnable examples}} \\ \cline{2-12} 
 & \multicolumn{2}{c|}{\multirow{2}{*}{\textbf{0\%}}} & \multicolumn{2}{c|}{\textbf{20\%}} & \multicolumn{2}{c|}{\textbf{40\%}} & \multicolumn{2}{c|}{\textbf{60\%}} & \multicolumn{2}{c|}{\textbf{80\%}} & \multirow{2}{*}{\textbf{100\%}} \\
 & \multicolumn{2}{c|}{} & \textbf{$\mathcal{D}_{u}+\mathcal{D}_{c}$} & \multicolumn{1}{c|}{\textbf{$\mathcal{D}_{c}$}} & \textbf{$\mathcal{D}_{u}+\mathcal{D}_{c}$} & \multicolumn{1}{c|}{\textbf{$\mathcal{D}_{c}$}} & \textbf{$\mathcal{D}_{u}+\mathcal{D}_{c}$} & \multicolumn{1}{c|}{\textbf{$\mathcal{D}_{c}$}} & \textbf{$\mathcal{D}_{u}+\mathcal{D}_{c}$} & \multicolumn{1}{c|}{\textbf{$\mathcal{D}_{c}$}} &  \\ \hline
\multicolumn{1}{c|}{$K=10$} & \multicolumn{2}{c|}{91.2} & 90.8 & \multicolumn{1}{c|}{91.0} & 90.0 & \multicolumn{1}{c|}{89.2} & 89.9 & \multicolumn{1}{c|}{87.2} & 88.3 & \multicolumn{1}{c|}{89.2} & 11.8 \\
\multicolumn{1}{c|}{$K=30$} & \multicolumn{2}{c|}{90.9} & 91.3 & \multicolumn{1}{c|}{91.2} & 90.2 & \multicolumn{1}{c|}{90.0} & 89.3 & \multicolumn{1}{c|}{90.1} & 89.2 & \multicolumn{1}{c|}{87.4} & 18.7 \\
\multicolumn{1}{c|}{$K=50$} & \multicolumn{2}{c|}{91.5} & 89.7 & \multicolumn{1}{c|}{90.9} & 89.2 & \multicolumn{1}{c|}{90.3} & 88.5 & \multicolumn{1}{c|}{89.3} & 86.9 & \multicolumn{1}{c|}{88.7} & 10.3 \\ \hline
\end{tabular}
}

\end{table}

\begin{figure}[!t]
    \begin{minipage}{0.51\linewidth}
        \includegraphics[width = \linewidth]{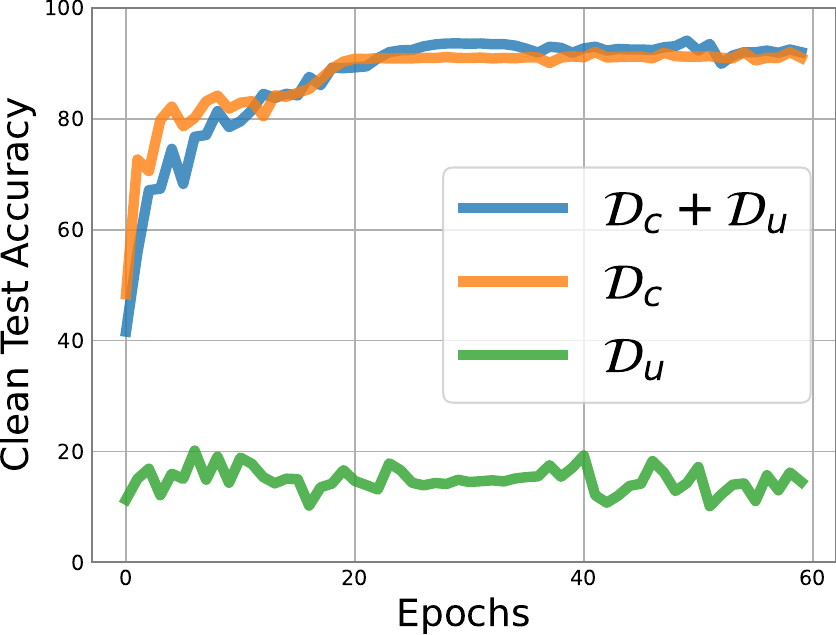}
    \end{minipage}
    \begin{minipage}{0.47\linewidth}
        \includegraphics[width = \linewidth]{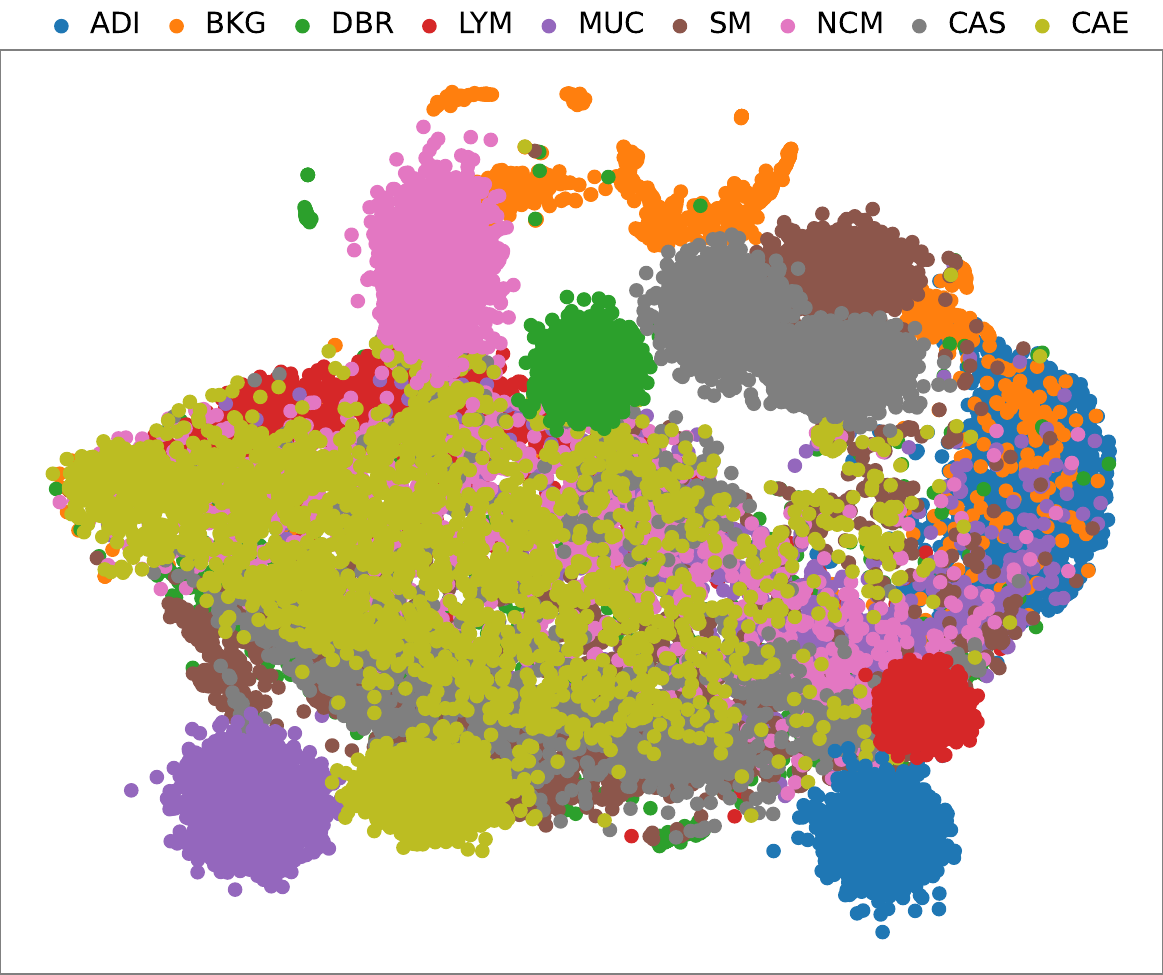}
    \end{minipage}
    \caption{\textbf{Left:} The learning curve trained on RN-18 on three different datasets. \textbf{Right: }The t-SNE results of the whole mixed data.}
    \label{fig: curve and tsne}
\end{figure}

In addition to the scenarios mentioned above, researchers may only be responsible for a portion of the data within the whole dataset. They extensively collect various classes of data and annotate them, then merge them into a complete dataset based on the labels. This spurred our investigation into the effectiveness of randomly selecting a subset of samples for perturbation. Specifically, we applied SALM to a selected percentage of the training data, leaving the remainder untouched and clean. We trained models on this mixed dataset of unlearnable and clean training data, $\mathcal{D}_u + \mathcal{D}_c$. For comparison, models are also trained on a completely clean dataset, denoted as $\mathcal{D}_c$.

The results in Table~\ref{tab: diff per} show a rapid decrease in effectiveness when less than 100\% of the data is selected for SALM application, whatever the $K$ is. Surprisingly, applying SALM to as much as 80\% of the data results in a negligible effect. Previous studies have shown that both error-minimizing noise and error-maximizing noise have similar limitations in DNNs~\cite{huang2021unlearnable,shan2020fawkes}.

To further illustrate this phenomenon, we take 80\% as an example and plot the learning curves of ResNet-18 in the following scenarios: 1) Models trained on 20\% clean data; 2) Models trained on 80\% data processed by SALM; 3) Models trained on a mixture of these two types of data. The results are shown in Figure~\ref{fig: curve and tsne} left. It is observed that data processed by SALM at 80\% remains unlearnable for the model, yet the mere 20\% of clean data enables the model to achieve excellent performance. The superior performance trained on mixed data arises solely from the clean data contained within. Furthermore, the visualization results in Figure~\ref{fig: curve and tsne} reveal that the clean data remains mixed in 2D space, whereas the data processed by SALM is almost linearly separable, indicating its unlearnability. Therefore, in the case of sample-wise combined datasets, our method also demonstrates outstanding performance and robust scalability. Overall, our method provides a way for every collaborating entity within large datasets to protect their interests and ensure each data donor's contribution will not be misused.

\clearpage
% \section{Visulization}
% \vspace{cm}
\begin{figure}[h]
    \centering
    \begin{minipage}[t]{0.4\textwidth}
        \includegraphics[width = \linewidth]{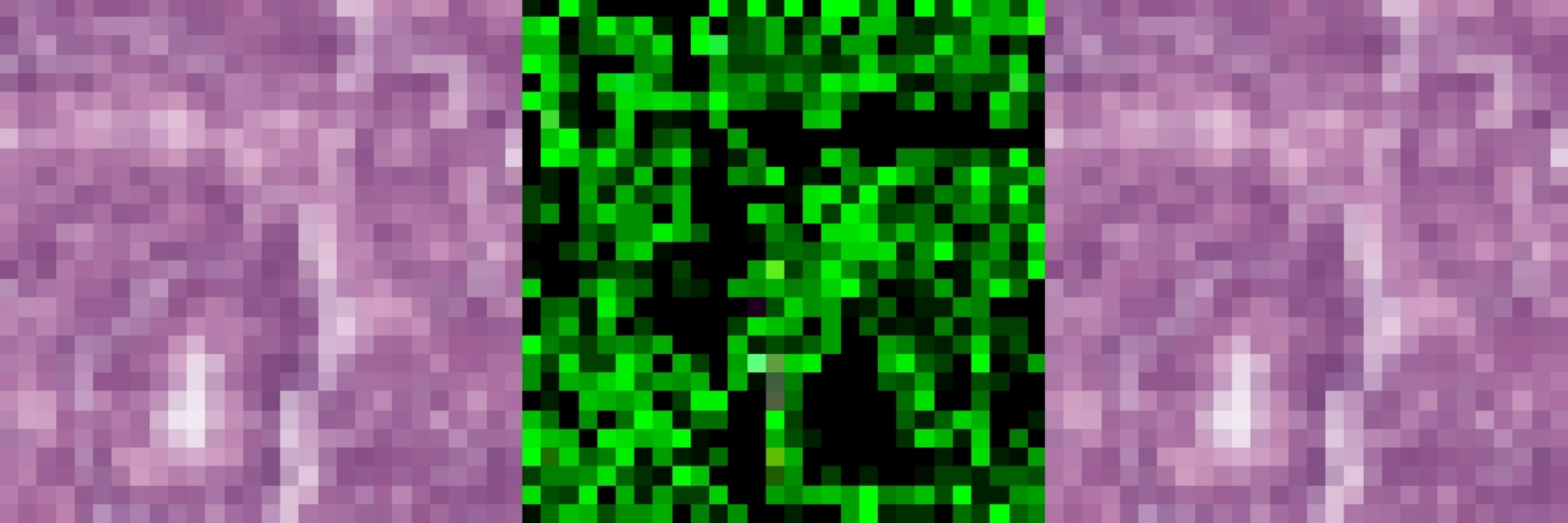}
    \end{minipage}
    \begin{minipage}[t]{0.4\textwidth}
        \includegraphics[width = \linewidth]{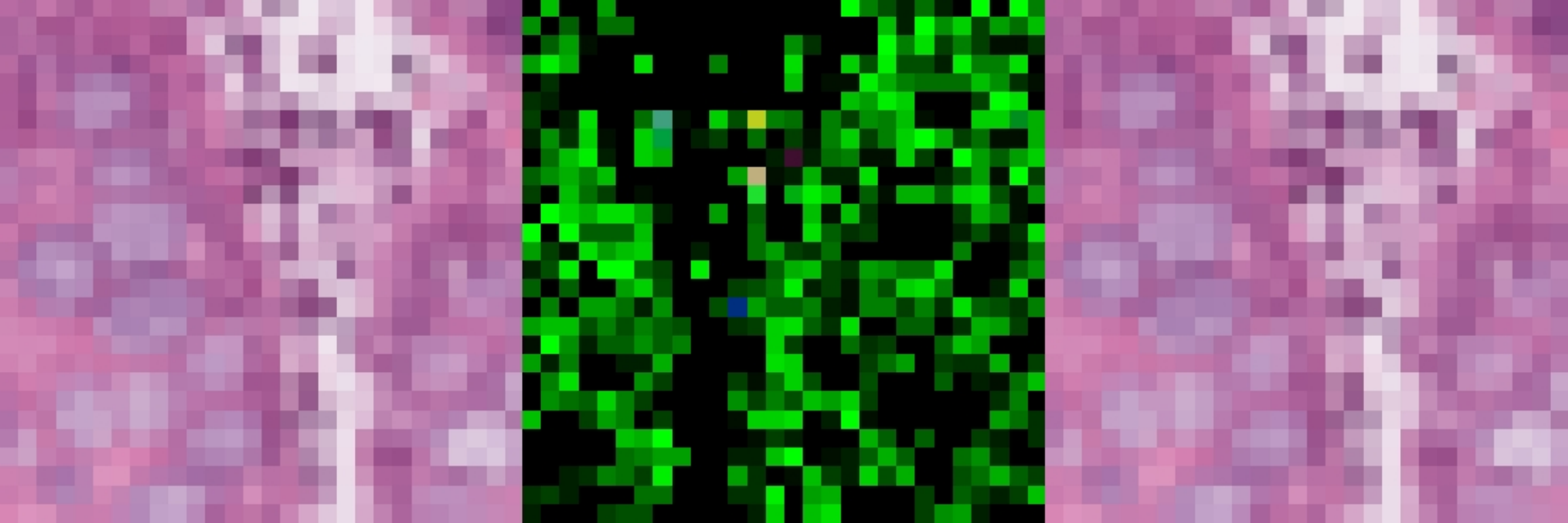}
    \end{minipage}
    \begin{minipage}[t]{0.4\textwidth}
        \includegraphics[width = \linewidth]{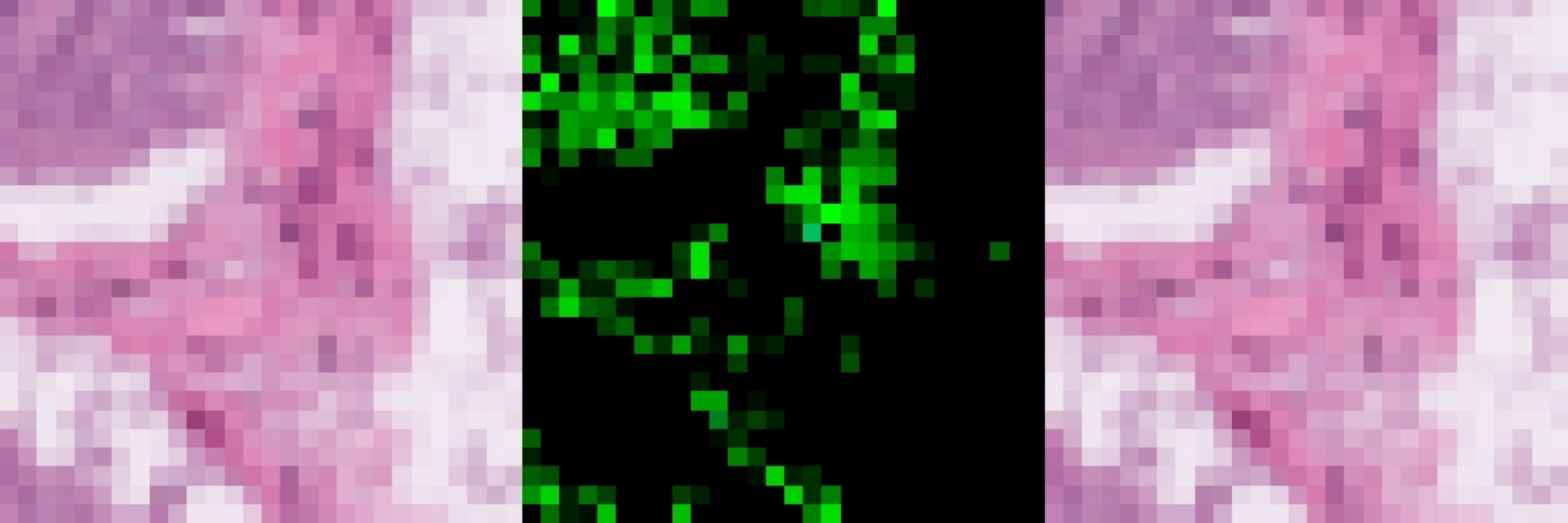}
    \end{minipage}
    \begin{minipage}[t]{0.4\textwidth}
        \includegraphics[width = \linewidth]{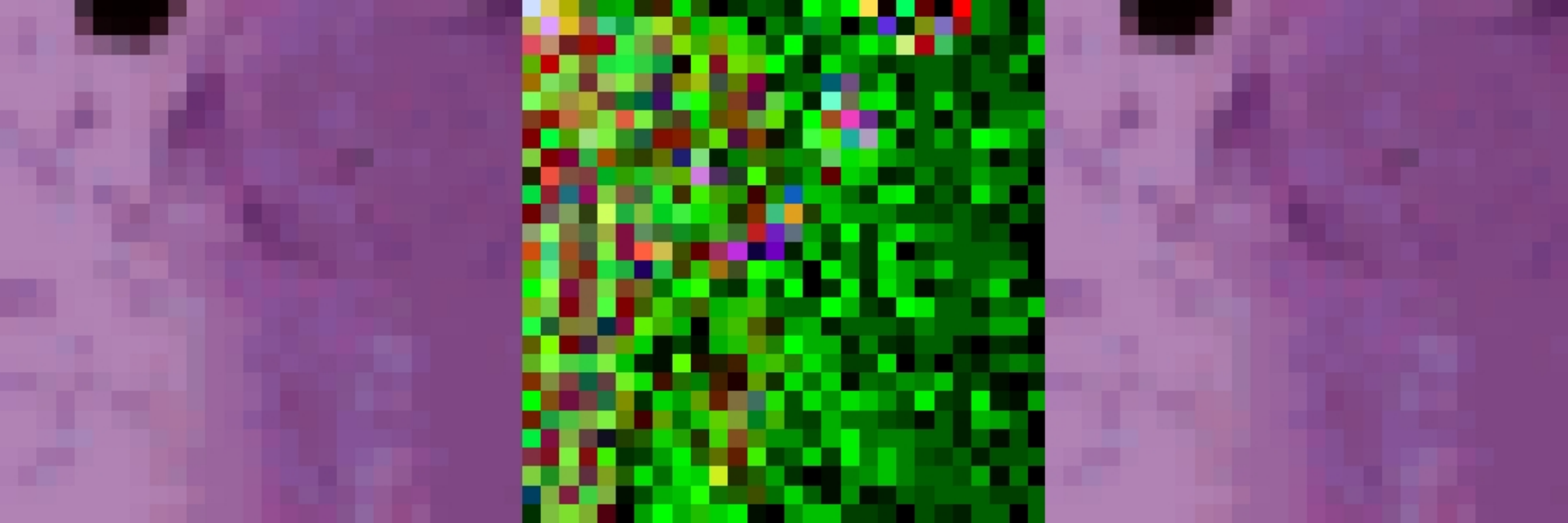}
    \end{minipage}
    \begin{minipage}[t]{0.4\textwidth}
        \includegraphics[width = \linewidth]{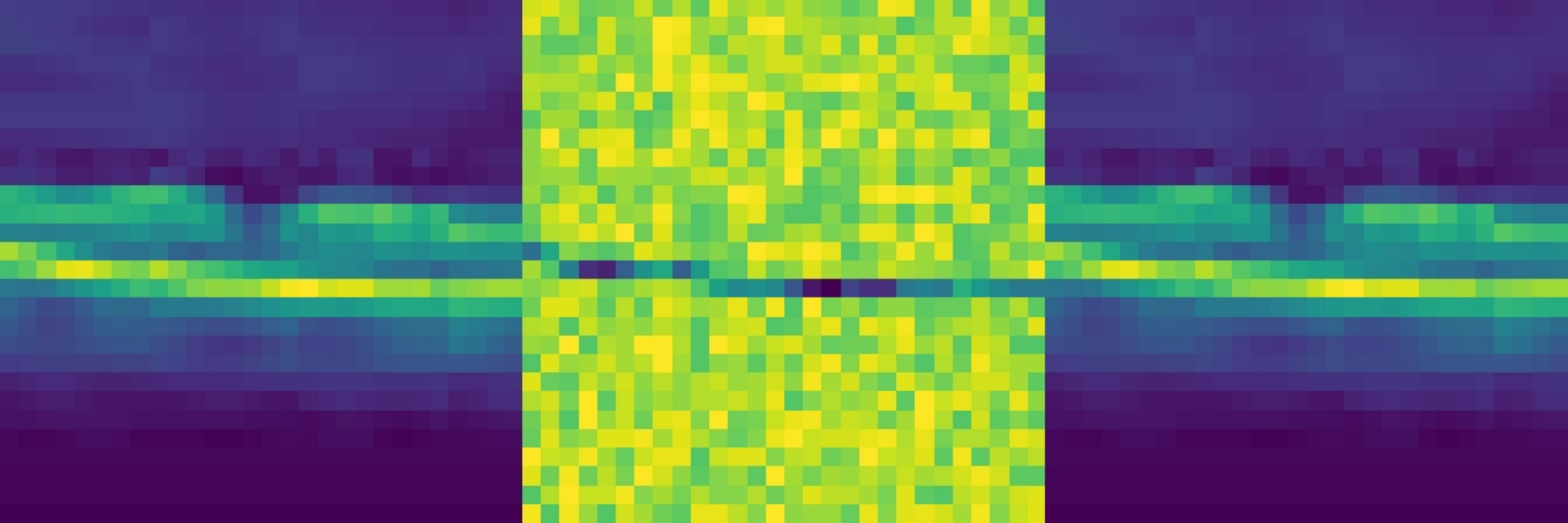}
    \end{minipage}
    \begin{minipage}[t]{0.4\textwidth}
        \includegraphics[width = \linewidth]{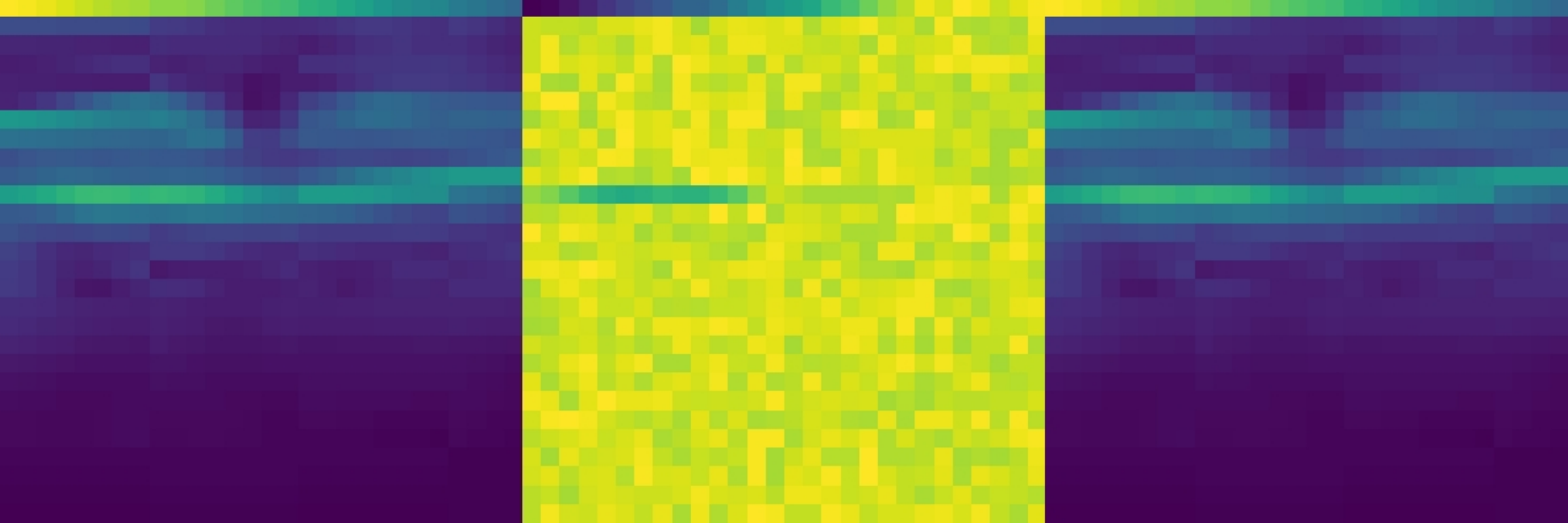}
    \end{minipage}
    \begin{minipage}[t]{0.4\textwidth}
        \includegraphics[width = \linewidth]{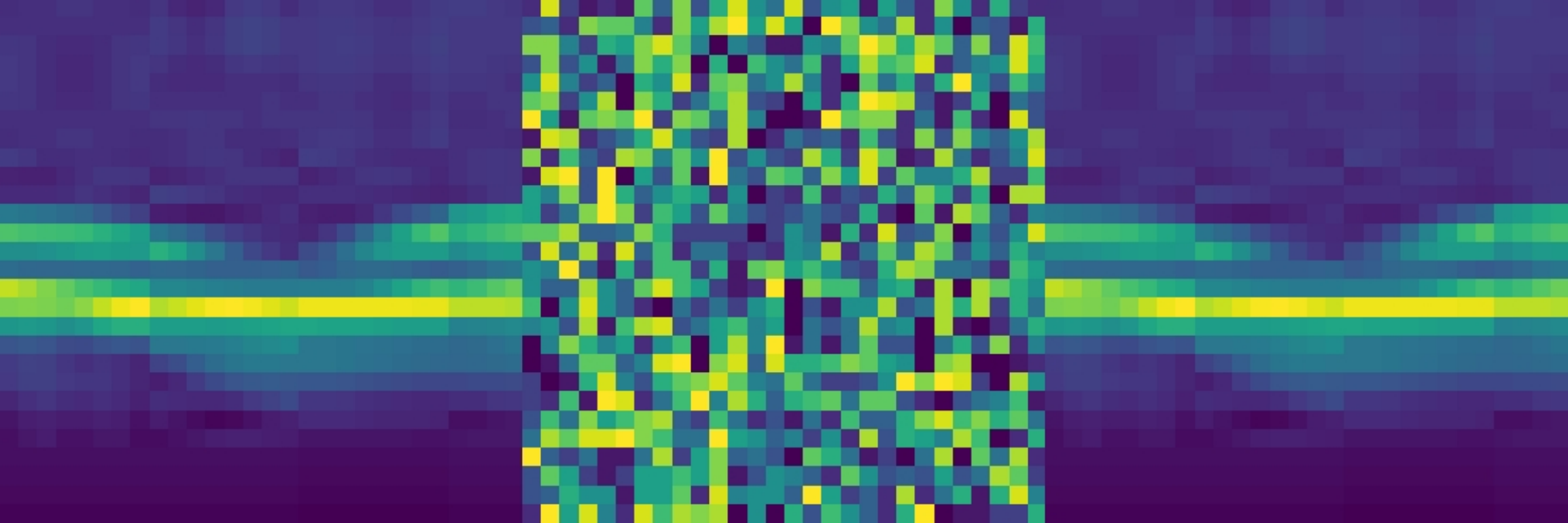}
    \end{minipage}
    \begin{minipage}[t]{0.4\textwidth}
        \includegraphics[width = \linewidth]{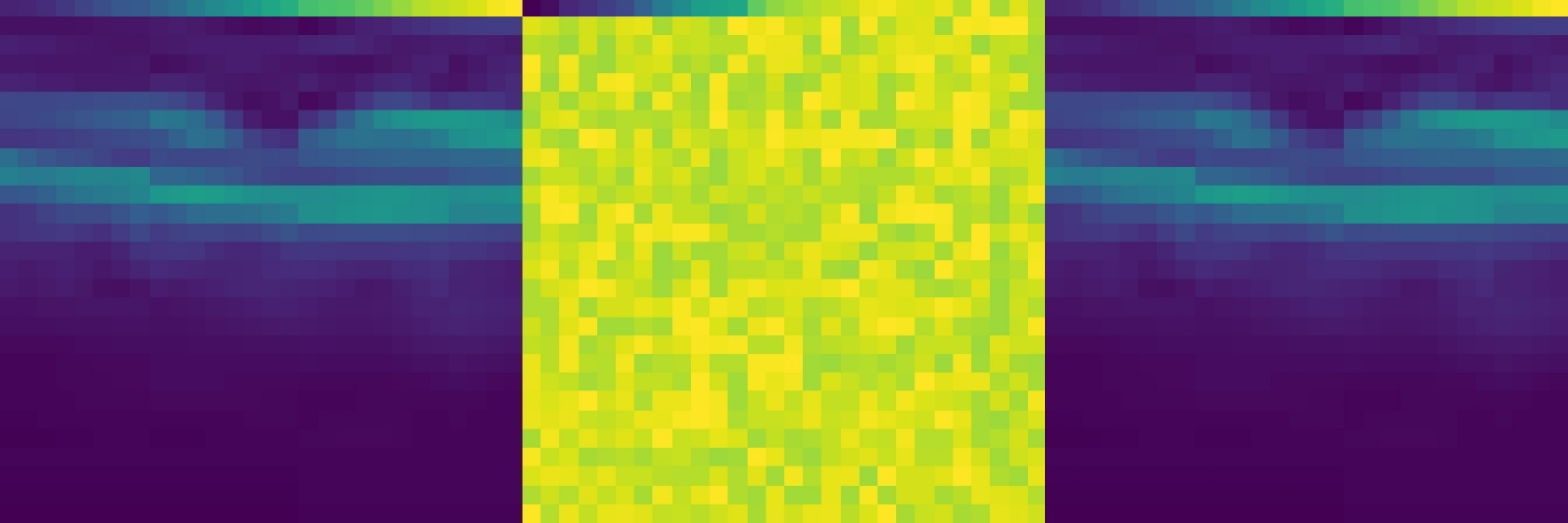}
    \end{minipage}
    \begin{minipage}[t]{0.4\textwidth}
        \includegraphics[width = \linewidth]{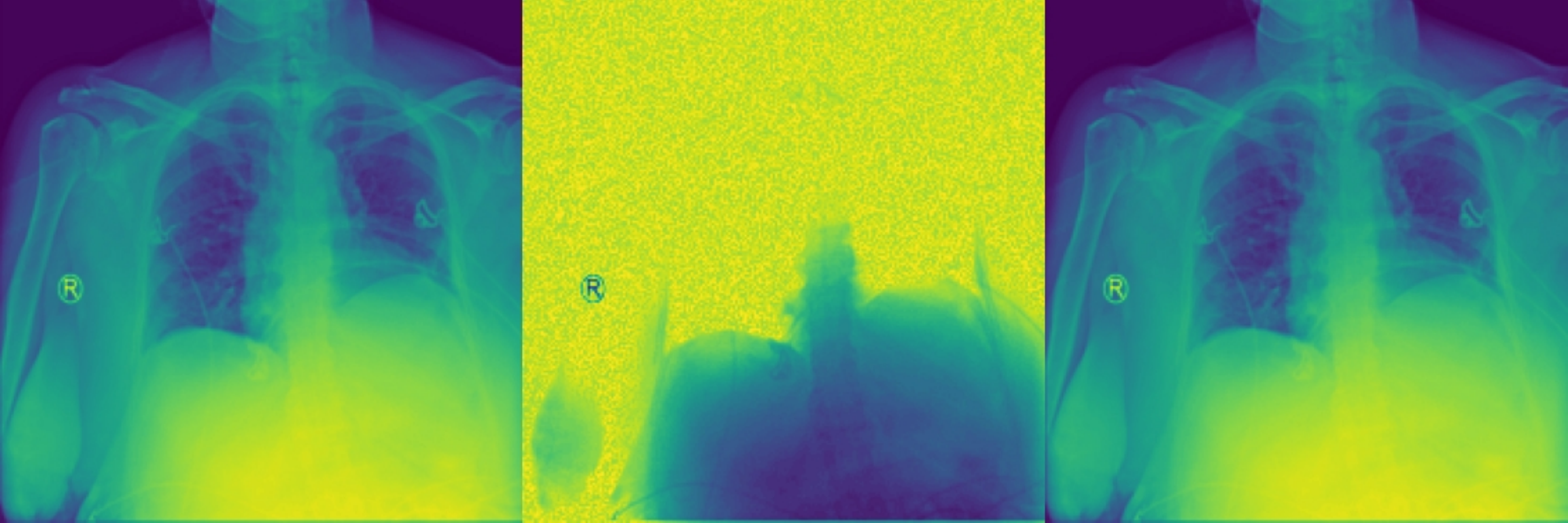}
    \end{minipage}
    \begin{minipage}[t]{0.4\textwidth}
        \includegraphics[width = \linewidth]{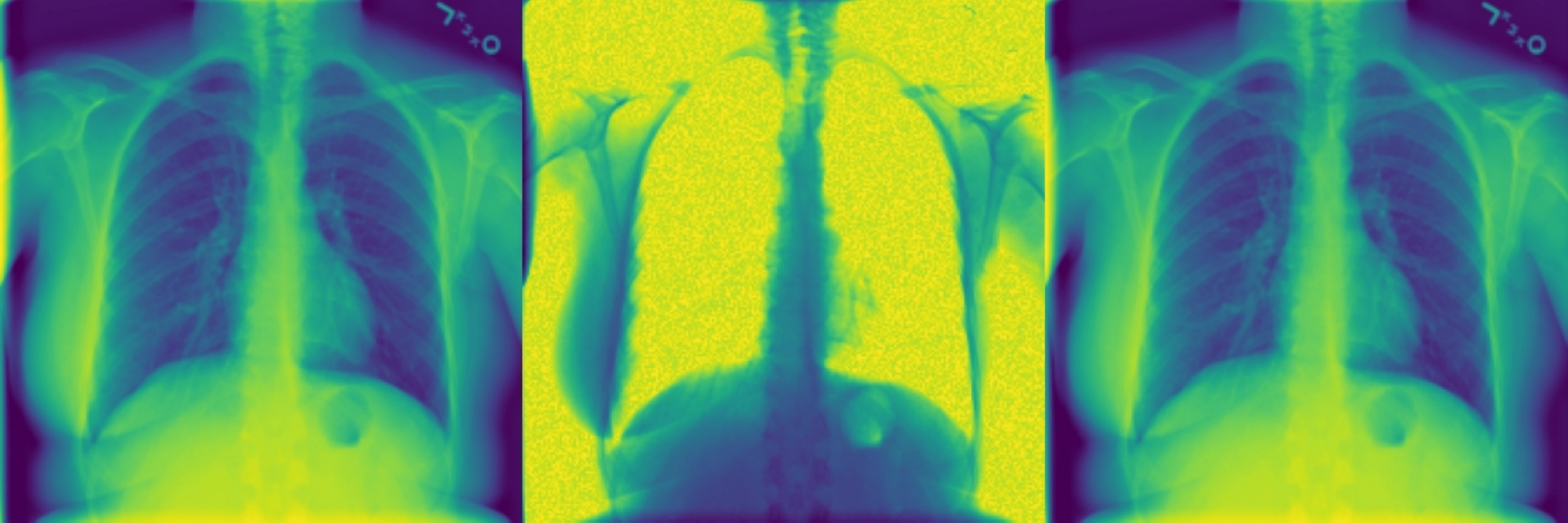}
    \end{minipage}
    \begin{minipage}[t]{0.4\textwidth}
        \includegraphics[width = \linewidth]{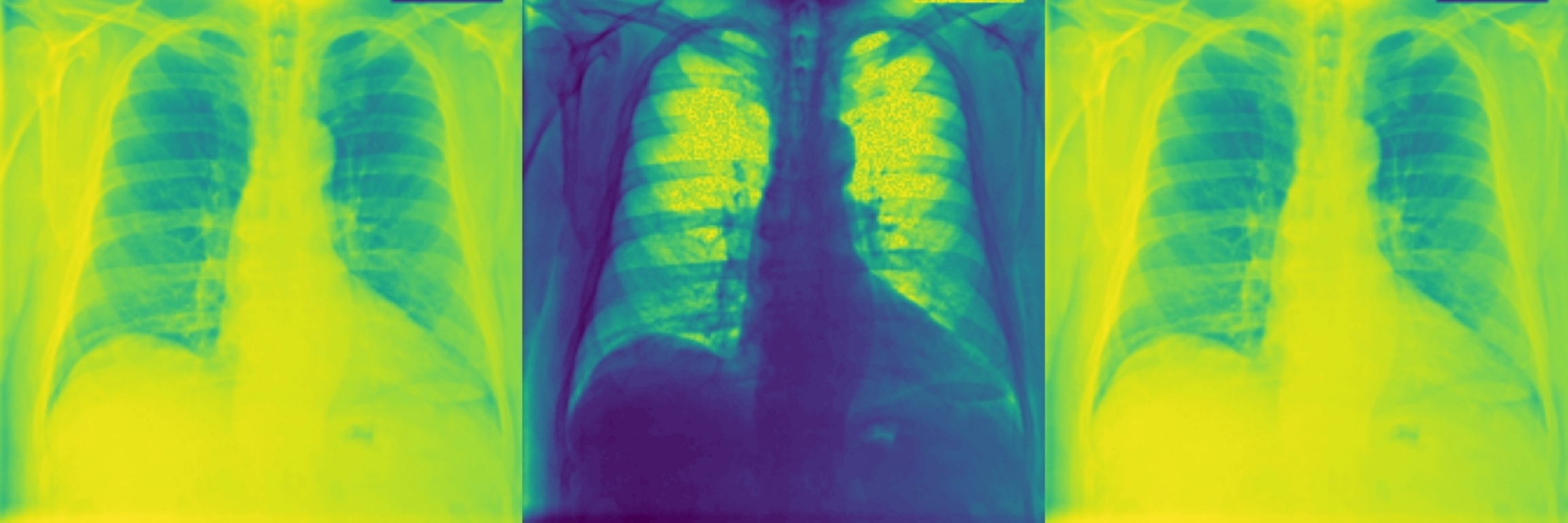}
    \end{minipage}
    \begin{minipage}[t]{0.4\textwidth}
        \includegraphics[width = \linewidth]{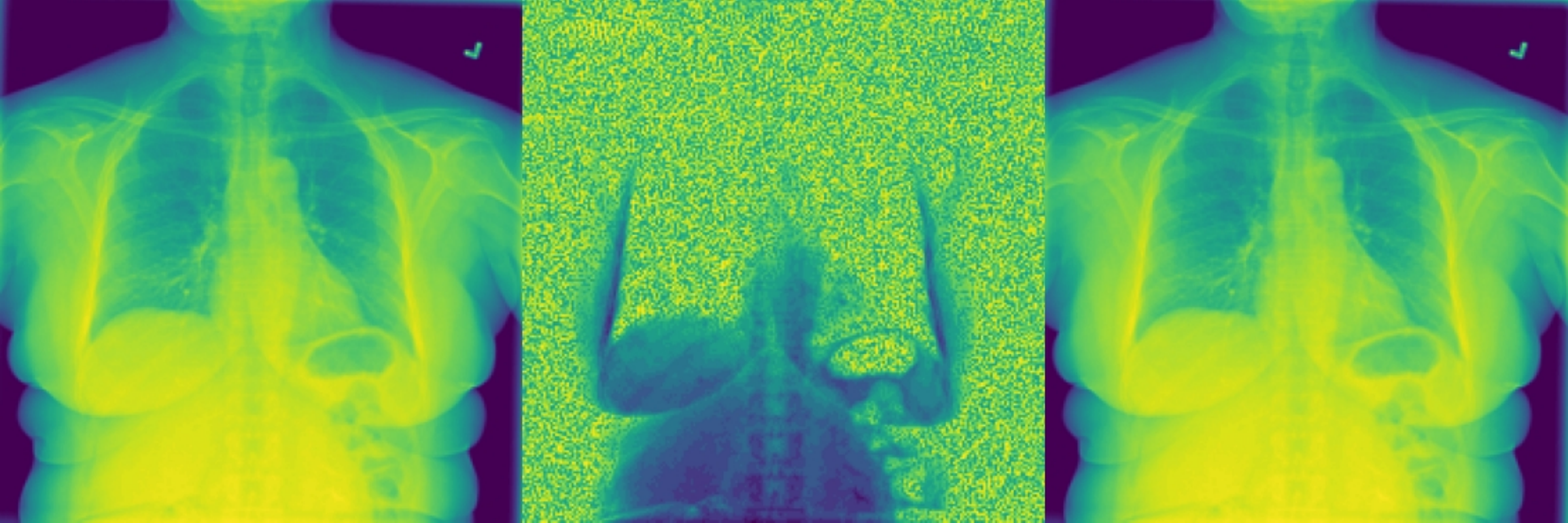}
    \end{minipage}
    \begin{minipage}[t]{0.4\textwidth}
        \includegraphics[width = \linewidth]{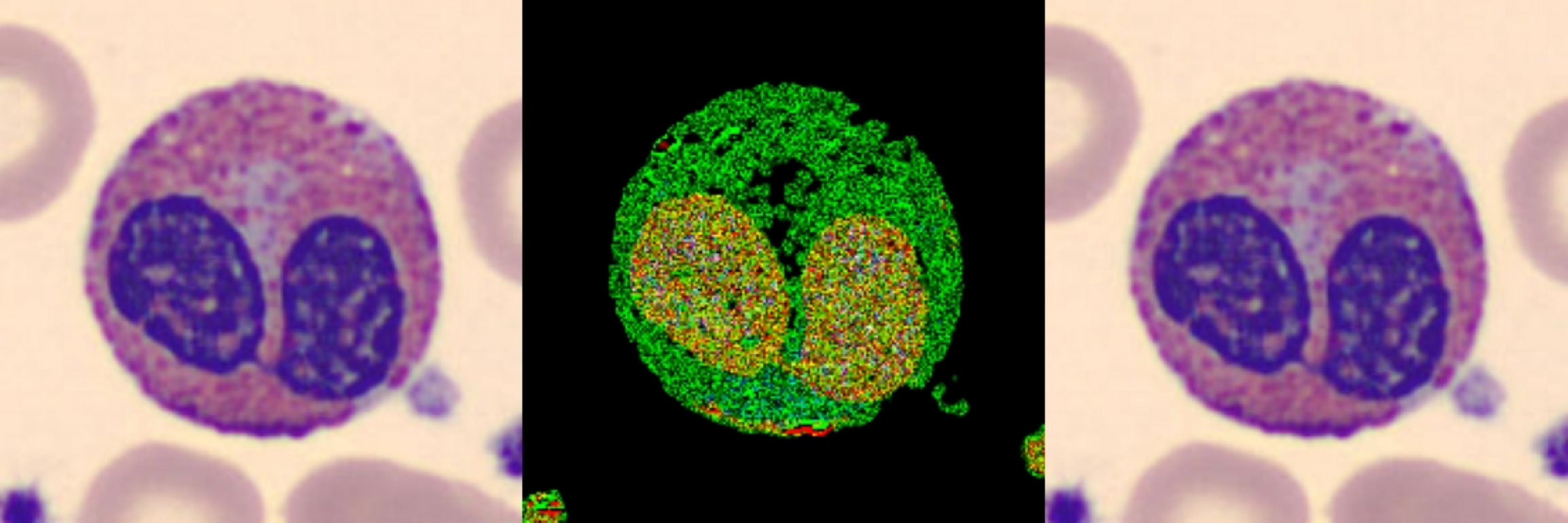}
    \end{minipage}
    \begin{minipage}[t]{0.4\textwidth}
        \includegraphics[width = \linewidth]{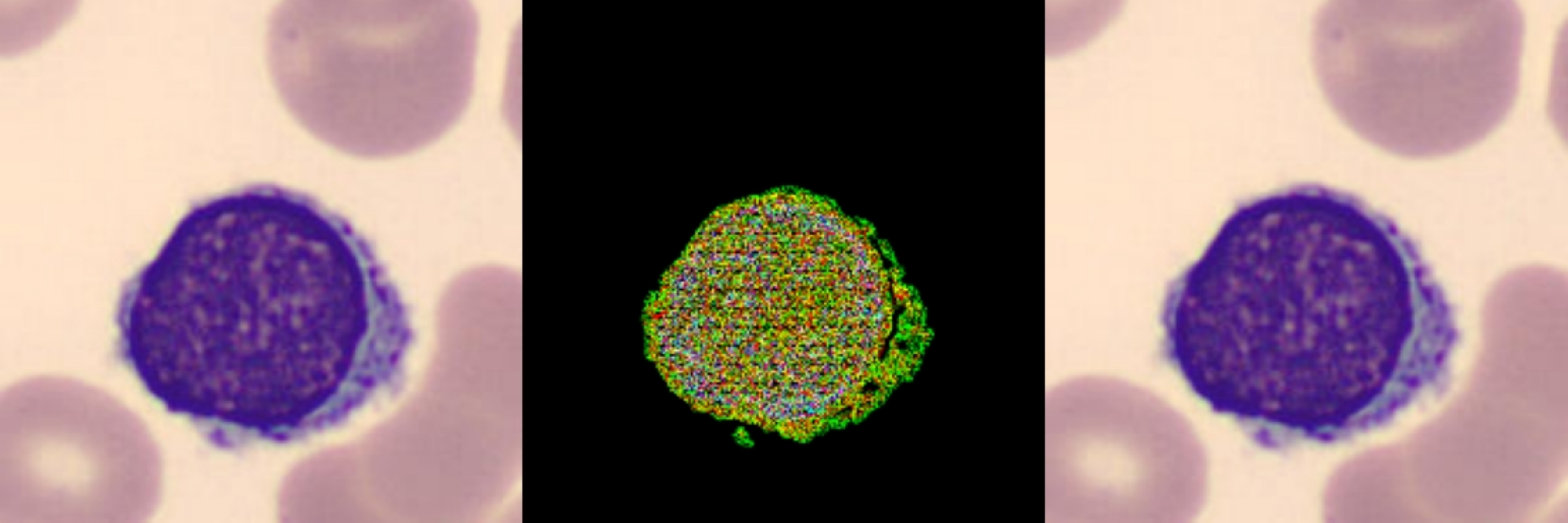}
    \end{minipage}
    \begin{minipage}[t]{0.4\textwidth}
        \includegraphics[width = \linewidth]{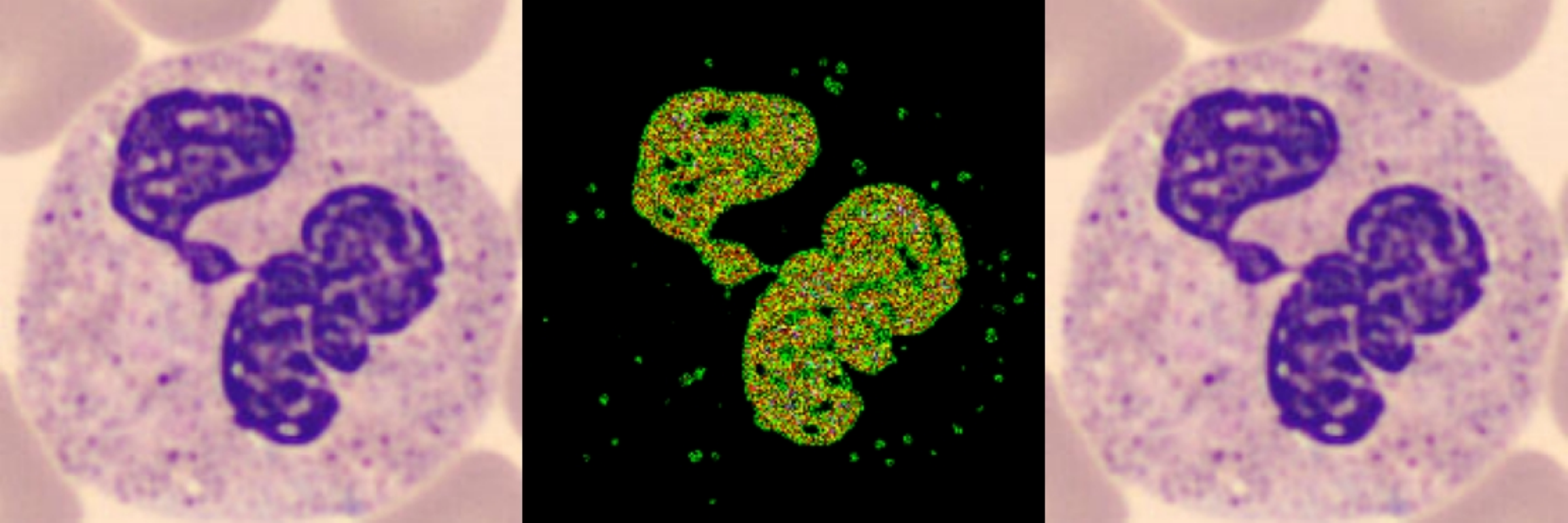}
    \end{minipage}
    \begin{minipage}[t]{0.4\textwidth}
        \includegraphics[width = \linewidth]{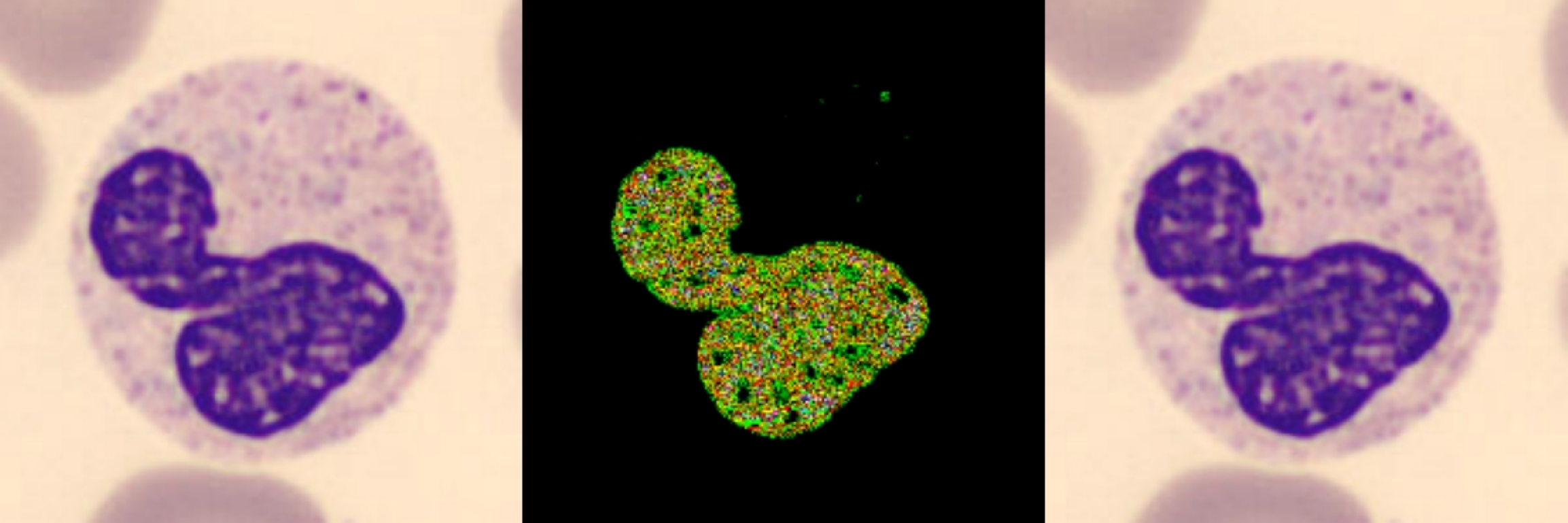}
    \end{minipage}

    \caption{Some visualization results of the origin image and the corresponding noise and protected image.}
    
\end{figure}

%%%%%%%%%%%%%%%%%%%%%%%%%%%%%%%%%%%%%%%%%%%%%%%%%%%%%%%%%%%%%%%%%%%%%%%%%%%%%%%
%%%%%%%%%%%%%%%%%%%%%%%%%%%%%%%%%%%%%%%%%%%%%%%%%%%%%%%%%%%%%%%%%%%%%%%%%%%%%%%

\end{document}